\documentclass[aps, 10pt, english, twoside, pra, nofootinbib, tightenlines, longbibliography, superscriptaddress]{revtex4-1}

\usepackage{document_config}

\begin{document}
    \title{Causal Compatibility Inequalities Admitting Quantum Violations in the Triangle Structure}
    \author{Thomas C. Fraser}
    \email{tfraser@perimeterinstitute.ca}
    \affiliation{Perimeter Institute for Theoretical Physics, Waterloo, Ontario, Canada, N2L 2Y5}
    \affiliation{University of Waterloo, Waterloo, Ontario, Canada, N2L 3G1}
    \author{Elie Wolfe}
    \affiliation{Perimeter Institute for Theoretical Physics, Waterloo, Ontario, Canada, N2L 2Y5}
    \date{\today}
    \begin{abstract}
        It has long been recognized that certain quantum correlations are incompatible with a particular assumption about classical causal structure. Given a causal structure of unknown classicality, the presence of such correlations certifies the nonclassical nature of the causal structure in a device-independent fashion. In structures where all parties share a common resource, these nonclassical correlations are also known as nonlocal correlations. Any constraint satisfied by all correlations which are classically compatible with a given causal structure defines a causal compatibility criterion. Such criteria were recently derived for the Triangle structure (E. Wolfe \textit{et al.}, \href[pdfnewwindow]{http://arxiv.org/abs/1609.00672}{arXiv:1609.00672}) in the form of polynomial inequalities, begging the question of whether any of those inequalities admit violation by quantum correlations. Numerical investigation suggests that they do not, and we further conjecture that the set of correlations admitted by the classical Triangle structure is equivalent to the set of correlations admitted by its quantum generalization whenever the three observable variables are binary. Our main contribution in this work, however, is the derivation of causal compatibility inequalities for the Triangle structure which do admit quantum violation. This provides a robust-to-noise witness of quantum correlations in the Triangle structure. We conclude by considering the possibility of quantum resources potentially qualitatively different from those known previously.
    \end{abstract}
    \maketitle

    \section{Introduction}
    \label{sec:introduction}
    In recent decades, the technical utility of quantum mechanics has become abundantly clear. In the realm of computation, quantum algorithms, such as Shor's algorithm~\cite{Shor_1997} and numerous others~\cite{Jordan_2016}, scale exponentially better than their classical counterparts. In the realm of secure communication, quantum protocols, a popular example being quantum key distribution~\cite{Bennett_2014}, are able to provide privacy even against hypothetical adversaries with unlimited computational power, a desideratum which classical protocols are unable to fulfill. Throughout history, numerous quantum phenomena which fail to be emulated by classical physics have been identified as resources for solving computational or communication problems~\cite{Neilsen_Chaung_2011}. Motivated by past successes, a primal objective of modern quantum information theory is to discover new situations wherein quantum mechanics offers an advantage, and to certify that the quantum advantage is genuine.

    From a foundational prospective, the most robust demonstrations of quantum phenomena with no classical emulation have involved Bell inequalities~\cite{Bell_1964,Brunner_2013}. Originally, Bell inequalities were derived as a way to show that no hidden variable theory could ever account for quantum mechanics; in this sense Bell inequalities are a response to the famous Einstein-Podolsky-Rosen paradox~\cite{EPR_Orig}. The enumeration of Bell inequalities has since become a widespread systematic method for demonstrating the nonclassicality of a given observation. More recently it has been appreciated that Bell inequalities can be understood as consequences of \term{causal inference}~\cite{Wood_2012}. Causal inference is concerned with classifying observations into those which can and cannot be explained by a hypothesized causal structure. The abstract nature of causal inference is responsible for its presence in numerous scientific fields including machine learning and biology~\cite{Pearl_2009,Pearl_2009_tr}. Causal compatibility inequalities, such as Bell and Instrumental inequalities \cite{pearl1995instrumental,bonet2001instrumental,evans2012graphical}, characterize the space of observations that are compatible with a hypothesized causal structure, albeit the characterization offered by practically derivable inequalities is often only an approximation. To derive the traditional Bell inequalities from causal inference one starts with a (classical) causal structure known as the Bell structure, as depicted in \cref{fig:bell_structure}. The fundamental Bell structure involves noncommunicating parties making measurements on some hidden shared resource $\la$, where the measurement outcomes ($A$ and $B$) are presumed to be stochastic functions of the local choices of measurement settings ($S_A$ and $S_B$) and the shared resource $\la$. Quantum nonclassicality in the Bell structure has been thoroughly studied since Bell's original work~\cite{Brunner_2013}. More complex structures, however, such as the correlation scenarios proposed by Fritz~\cite{Fritz_2012,Fritz_2014}, are much less understood. Here we investigate one particular correlation scenario named the Triangle structure (\cref{fig:triangle_structure}).

    The Triangle structure (\cref{fig:triangle_structure}) is a causal structure comprised of three parties labeled $A, B,$ and $C$ arranged in a triangular configuration while pairwise sharing hidden (latent) variables $X, Y,$ and $Z$. It has been extensively studied previously (see, e.g.,~\cite[Fig. 1]{Steudel_2010},~\cite[Fig. 6]{Chaves_2014},~\cite[Fig. 8]{Branciard_2012},~\cite[Fig. 8, App. E]{Henson_2014},~\cite[Fig. 3]{Fritz_2012},~\cite[Fig. 4]{Weilenmann_2016}, and~\cite[Fig. 1]{Inflation}). An overview of some milestone results is provided in \cref{sec:triangle_structure}. Identifying causal compatibility inequalities for this configuration has been seen as particularly challenging~\cite{Branciard_2012}. Further identifying causal compatibility inequalities of such high resolution that such inequalities can be violated by quantum-accessible distributions has remained out of reach for the Triangle structure.

    This work finds causal compatibility inequalities for the Triangle structure that are known to be violated by quantum-accessible distributions. This accomplishment was made possible through the combination of two previous developments: first, the insight of \citet{Fritz_2012} regarding the ability to reinterpret the Bell structure as a portion of the Triangle structure, and second, the framework for solving causal inference problems developed by~\citet{Inflation} called \term{The Inflation Technique}. Ultimately, this work serves as a validation that the Inflation Technique is efficient and sensitive enough at low orders to offer insights into quantum nonclassicality \cite{Navascues_2017}. Moreover, these inequalities offer an avenue for recognizing previously unknown forms of nonclassicality. The authors' attempts to find such novel resources were met with only partial success, suggesting the need for both conceptual refinements and future exploration.

    The first half of the manuscript, namely \Cref{sec:causal_compatibility,sec:bell_structure,sec:triangle_structure,sec:fritz_distribution}, is entirely a review of previous works. \Cref{sec:causal_compatibility} recalls important notions from causal inference theory and sets up the notation to be used. \Cref{sec:bell_structure} offers a summary of the popular Bell structure and associated inequalities. \Cref{sec:triangle_structure} discusses the Triangle structure and provides an overview of existing research, identifying its stark differences from the Bell structure and motivating why the Triangle structure is worth studying. \Cref{sec:fritz_distribution} defines and discusses a singularly quantum correlation first conceived of by \citet{Fritz_2012}, which we term the \term{Fritz distribution}. In the same work~\cite{Fritz_2012}, the Fritz distribution was proven to be nonclassical without the use of inequalities.

    The second half of the manuscript, namely \Cref{sec:preliminary_research,sec:found_inequalities,sec:optimizations,sec:fritz_problem_revisit}, presents the main contributions made by this research project. Specifically, in \cref{sec:found_inequalities} we improve upon the results of \citet{Fritz_2012} by offering a direct proof of the incompatibility of the Fritz distribution using inequalities generated by the Inflation Technique~\cite{Inflation}. A sample of such inequalities is presented in \cref{sec:found_inequalities}, specifically \cref{eq:ww_ineq}, \cref{eq:web_inequality}, and \cref{eq:symmetric_web_inequality}. Aside from confirming the utility of the Inflation Technique, this paper explores the importance of having derived these inequalities. First, an inequality-based proof has the advantage of being robust to experimental noise. In \cref{sec:violations_noise} the Fritz distribution is subjected to noise in order to measure the robustness of the derived inequalities. Second, we numerically optimize our derived inequalities over quantum-accessible distributions (using qubits) in an effort to find the maximum violations achievable by quantum theory. The culmination of these analyses naturally prompts a discussion, found in \Cref{sec:fritz_problem_revisit}, regarding the fundamental problem of recognizing and classifying nonclassicality in the Triangle structure. \Cref{sec:conclusions} concludes.

    \Cref{sec:inflation_technique_summary} briefly summaries the Inflation Technique in the specific context of this work. Although the summary presented in \Cref{sec:inflation_technique_summary} is designed to be self-standing, a much more pedagogical introduction is offered by the original work~\cite{Inflation}. \Cref{sec:deriving_inequalities} demonstrates how the Inflation Technique was used to derive the causal compatibility inequalities for the Triangle structure which admit violation by quantum-accessible distributions.

    \section{Causal Compatibility}
    \label{sec:causal_compatibility}
    The task of causal inference is to determine the set of potentially observable probability distributions compatible with some hypothesis about causal relationships~\cite{Pearl_2009}. If an observed distribution can be explained by the hypothesized causal mechanism, then the distribution is said to be \textit{compatible} with said causal mechanism. In order to define compatibility rigorously, we first need to formally define the notion of a causal hypothesis.

    A hypothesis of causal mechanism is formally referred to as a \term{causal structure} and can be represented as a \term{directed acyclic graph}. A directed graph $\graph$ is an ordered tuple $\br{\nodes, \edges}$ of respectively \textit{nodes} and \textit{edges} where each edge $e \in \edges$ connects a \textit{pair} of nodes $n, m \in \nodes$ with a directed arrow $e = \bc{n \to m}$. A directed graph is \textit{acyclic} if there are no paths following the directions of the edges starting from and returning to the same node. The nodes $\nodes$ of a causal structure represent random variables while the edges $\edges$ represent a casual influence from one variable to another pursuant to the prescribed direction.

    Henceforth, we will utilize a number of familiar notions from graph theory and denote them accordingly. The \term{parents of a node} $n \in \nodes$ are all nodes which point directly into $n$, i.e. $\Pa[\graph]{n} \defined \bc{m \mid m \to n}$. Similarly defined are the \term{children of a node} $\Ch[\graph]{n} \defined \bc{m \mid n \to m}$. Recursively defined are the \term{ancestors of a node} $\An[\graph]{n} \defined \bigcup_{i\in\mathbb{N}} \Pa[\graph][i]{n}$ where $\Pa[\graph][i]{n} \defined \Pa[\graph]{\Pa[\graph][i-1]{n}}$ and $\Pa[\graph][0]{n} = n$ and the \term{descendants of a node} $\De[\graph]{n} \defined \bigcup_{i\in\mathbb{N}} \Ch[\graph][i]{n}$ where $\Ch[\graph][i]{n} \defined \Ch[\graph]{\Ch[\graph][i-1]{n}}$ and $\Ch[\graph][0]{n} = n$. Finally, we extend this notation to a subset of nodes $N \subseteq \nodes$ by performing a union over elements. As an example, the parents of the nodes $N \subseteq \nodes$ are denoted $\Pa[\graph]{N} = \bigcup_{n \in N} \Pa[\graph]{n}$.

    Let us now formalize the notion of compatibility between a causal structure $\graph$ and a probability distribution $\prob[\nodes]$ defined over the nodes of $\graph$. A causal structure $\graph$ hypothesizes that each variable $n \in \nodes$ is only directly influenced by its parents $\Pa[\graph]{n}$. Therefore, if each variable is conditioned on its parentage, the probability distribution $\prob[\nodes]$ should factorize accordingly:
    \[ \prob[\nodes] = \prod_{n \in \nodes} \prob[n \mid \Pa[\graph]{n}] = \prob[n_1 \mid \Pa[\graph]{n_1}] \times \cdots \times \prob[n_k \mid \Pa[\graph]{n_k}] \eq \label{eq:compatibility_as_factorization}\]

    If a given distribution $\prob[\nodes]$ defined over \textit{all} of the nodes $\nodes$ of $\graph$ satisfies \cref{eq:compatibility_as_factorization}, then $\prob[\nodes]$ is said to be \term{compatible with $\graph$}. The conditional distributions in \cref{eq:compatibility_as_factorization} (i.e. $\{\prob[n \mid \Pa[\graph]{n}] \mid n \in \nodes\}$) are referred to as a set of \term{causal parameters for $\graph$}. If a distribution $\prob[\nodes]$ cannot be factorized according to \cref{eq:compatibility_as_factorization}, $\prob[\nodes]$ is said to be \term{incompatible} with $\graph$. When one is specified with a joint distribution $\prob[\nodes]$ defined over all nodes of a causal structure $\graph$, it is possible to completely determine whether or not $\prob[\nodes]$ is compatible with $\graph$ by computing the causal parameters induced by $\prob[\nodes]$ and checking the equality of \cref{eq:compatibility_as_factorization}. A challenge, however, is presented when one is supplied with a partial observation, i.e. a joint distribution $\prob[\nodes_{O}]$ where $\nodes_{O} \subset \nodes$ is some subset of variables referred to as the \term{observable nodes} $\nodes_O$. In such cases, $\prob[\nodes_{O}]$ does not induce a unique set of causal parameters for $\graph$ and \cref{eq:compatibility_as_factorization} can not be verified by direction calculation. Instead, compatibility between $\graph$ and $\prob[\nodes_{O}]$ depends on the existence or nonexistence of a set of causal parameters for $\graph$ such that $\prob[\nodes_{O}] = \sum_{n \not \in \nodes_{O}} \prob[\nodes]$ where $\prob[\nodes]$ is again given by \cref{eq:compatibility_as_factorization}. The complementary, unobservable nodes are termed \term{latent nodes} $\nodes_L = \nodes \setminus \nodes_{O}$, and should be understood as hidden random variables that are unknowable either by some fundamental process or cannot be measured due to other limitations.

    There are several approaches to tackling the compatibility problem when dealing with latent variables; there are two common approaches worth mentioning here. The first is to recognize that many equality constraints are implied by the causal structure, including conditional independence relations and so-called Verma constraints among others; see Refs. \cite{evans2015margins,evans2017markov} for thorough treatments. The failure to satisfy an equality constraint immediately disqualifies $\prob[\nodes_{O}]$ from being compatible with $\graph$. Equality constraints are easily derived given a causal structure, and checking equality-constraint satisfaction is the minimalistic algorithm which powers the overwhelming majority of practical causal inference hypothesis testing in the fields of machine learning and artificial intelligence. In quantum theory, however, we require strong, more sensitive, causal inference techniques. This is because the equality constraints satisfied by compatible classical correlations are also all satisfied by quantum correlations \cite{Henson_2014}.
    Our focus, therefore, in on deriving \emph{inequality} constraints (over $\prob[\nodes_{O}]$ and its marginals) implied by a causal structure, which we term \term{causal compatibility inequalities}\footnote{We refer to these inequalities as ``causal compatibility inequalities'' instead of Bell inequalities for two reasons. Firstly, ``Bell inequalities'' usually are associated specifically with the Bell structure. Secondly, the inequalities derived in this work are fundamentally distinct from a typical Bell inequality in that these inequalities are \textit{polynomial} over $\prob[\nodes_{O}]$ instead of \textit{linear}.}. For some causal structures, the equality constraints associated with it are sufficient to perfectly characterize the distributions genuinely compatible with it; for others, however, inequality constraints are also important. Causal structures for which inequality constraints are relevant have been termed \textit{interesting}~\cite{Henson_2014}, and such structures include the Instrumental structure, the Bell structure, and the Triangle structure studied here, among infinitely many others.
    Herein we use the Inflation Technique~\cite{Inflation} to find causal compatibility inequalities; \Cref{sec:inflation_technique_summary} discusses the Inflation Technique as applied in this work.

    If a probability distribution $\prob[\nodes_{O}]$ happens to \textit{violate} any causal compatibility inequality, then that distribution is deemed \textit{incompatible}.  Conversely, a singular inequality can only be used to prove that a given distribution is \textit{incompatible}; a single inequality cannot certify compatibility. A \textit{complete characterization} of compatibility consists of a complete set of all valid causal compatibility inequalities such that satisfaction of the entire set certifies compatibility. Currently however, it is unknown how to obtain a complete characterization for all causal structures, including the Triangle structure.

    From the perspective of identifying quantum nonclassicality, a causal structure $\graph$ adopts the role of a classical hypothesis. Therefore, nonclassicality becomes synonymous with incompatibility: if a distribution $\prob[\nodes_{O}]$ is incompatible, then it is nonclassical. Henceforth, we will use these two terms interchangeably. From a resource standpoint, if the nonclassicality of $\prob[\nodes_{O}]$ with $\graph$ can be witnessed by an inequality $I$, but nevertheless $\prob[\nodes_{O}]$ can be implemented using quantum states and measurements while otherwise respecting the causal relations of $\graph$, then the causal compatibility inequality $I$ represents a task or game where quantum resources outperform classical resources relative to $\graph$.

    \section{Bell Structure}
    \label{sec:bell_structure}
    This section aims to define the Bell causal structure and to review some of the traditional witnesses used to assess the classicality (or lack thereof) of distributions relative to it. The purpose of this section is to equip readers with the pertinent background and also to draw comparisons between the advancements made toward understanding nonclassicality of the Bell structure versus analogous results obtained for Triangle structure later in this work.

\begin{figure}
    \begin{nscenter}
        \begin{minipage}[t]{.48\textwidth}
            \centering
            \scalebox{1.0}{\includegraphics{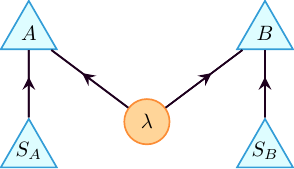}}
            \caption{The Bell structure consisting of two observers $A, B$ together with measurement settings $S_{A}$ and $S_{B}$ respectively. The shared latent variable is labeled $\la$.}
            \label{fig:bell_structure}
        \end{minipage}\hspace{0.04\textwidth}%
        \begin{minipage}[t]{.48\textwidth}
            \centering
            \scalebox{1.0}{\includegraphics{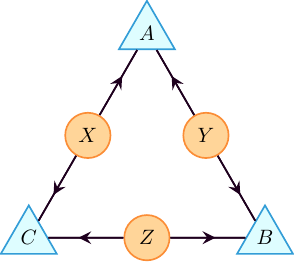}}
            \caption{The Triangle structure consisting of three observable variables $A,B,C$ and three latent variables $X, Y, Z$.}
            \label{fig:triangle_structure}
        \end{minipage}
    \end{nscenter}
    \end{figure}

    The bipartite Bell structure (\cref{fig:bell_structure}) refers to an iconic causal structure involving two distant parties who observe the outcomes of local measurements as random variables $A,B$ determined by their individual measurement settings $S_A,S_B$, and where the parties are also presumed to be commonly informed by some shared latent resource $\la$~\cite{Brunner_2013}. The observed correlations naturally form a conditional probability distribution $\prob[AB|S_AS_B]$. Subject to the notions of compatibility presented in \cref{sec:causal_compatibility}, $\prob[AB|S_AS_B]$ is compatible with \cref{fig:bell_structure} if and only if there exists some distribution $\prob[\la]$ and causal parameters $\prob[A|S_A,\la]$, $\prob[B|S_B,\la]$ such that\footnote{Here the summation $\sum_{\la}$ is used to denote a statistical marginalization over the latent variable $\la$ with unspecified support.}

    \begin{align}\label{eq:bellcompat}
     \prob[AB|S_AS_B] \text{ is classically compatible with \cref{fig:bell_structure}} \quad\iff \quad \prob[AB|S_AS_B] = \sum_{\la} \prob[A|S_A,\la]\prob[B|S_B,\la]\prob[\la]\,.
    \end{align}

    Any observed distribution $\prob[AB|S_AS_B]$ which fails to be explained by the classical causal hypothesis of the Bell structure as defined by \cref{eq:bellcompat} is appropriately termed nonclassical. Often, distributions incompatible with the Bell structure are referred to as nonlocal because the Bell structure markedly lacks causal influence from one party to another. In particular,~\citet{Bell_1964} demonstrated that there exists distributions that are nonclassical yet are attainable via local measurements on a shared quantum resource.

    Let us now contrast the classical definition of compatibility with its quantum generalization. Quantum correlation arise from a quantumified version of the causal structure, in which latent variables are replaced with quantum systems. Concretely, a distribution is accessible with quantum resources and measurements for the quantumified Bell structure if and only if there exists a bipartite quantum state $\rho_{AB}$ of arbitrary dimension, as well as Hilbert-space-localized measurement sets $M_{A|S_A}, M_{B|S_B}$\footnote{Note that a measurement set $M_{A} = \bc{M_{A}^{1}, M_{A}^{2}, \ldots, M_{A}^{k}}$ services a shorthand notation in the following sense, $\prob[A] = \Tr\bs{\rho M_{A}} \implies \prob[A]\br{a} = \Tr\bs{\rho M_{A}^{a}}$.} with indices conditional upon the measurement settings, such that
    \begin{align}\label{eq:bellQcompat}
    \prob[AB|S_AS_B] \text{ is a quantum realization of \cref{fig:bell_structure}} \quad\iff\quad\prob[AB|S_AS_B] = \Tr\bs{\rho_{AB}M_{A|S_A}\otimes M_{B|S_B}}\,.
    \end{align}
    The nonclassicality of quantum distributions for the Bell structure can be demonstrated through the use of Bell inequalities which constrain the correlations between binary variables $A$ and $B$ for classically compatible distributions. A notable example is the Clauser-Horne-Shimony-Holt (CHSH) inequality~\cite{CHSH_Original}.
    \[ \ba{AB | S_{A} = 0, S_{B} = 0} + \ba{AB | S_{A} = 0, S_{B} = 1} + \ba{AB | S_{A} = 1, S_{B} = 0} - \ba{AB | S_{A} = 1, S_{B} = 1} \leq 2 \eq \label[ineq]{eq:CHSH_ineq_orig} \]
    Nontrivial Bell inequalities such as the CHSH inequality are capable of witnessing the nonclassical nature of quantum distributions; the inequalities presented in \cref{sec:found_inequalities} are also of this type. Equality constraints, as previously mentioned, never have that sort of high-resolution discernment sensitivity.

    \section{Triangle structure}
    \label{sec:triangle_structure}

    As was mentioned in \cref{sec:introduction,sec:causal_compatibility}, the \term{Triangle structure} (\cref{fig:triangle_structure}) is a causal structure $\graph$ consisting of $3$ observable variables $A, B, C$ arranged in a triangular configuration while pairwise sharing latent variables $X, Y, Z$.

    Following the definition of causal compatibility from \cref{sec:causal_compatibility}, a distribution $\prob[\nodes_{O}] = \prob[ABC]$ is compatible with the Triangle structure if and only if there exists a choice of causal parameters $\bc{\prob[A|X,Y],\prob[B|Y,Z],\prob[C|Z,X],\prob[X],\prob[Y],\prob[Z]}$ such that $\prob[ABC]$ is a marginalization of $\prob[ABCXYZ]$ over $X, Y, Z$\footnote{Here the summation $\sum_{X,Y,Z}$ is used to denote a statistical marginalization over the latent variables $X,Y,Z$ with unspecified support.}, i.e.
    \[ \prob[ABC] \text{ is classically compatible with \cref{fig:triangle_structure}} \quad\iff\quad\prob[ABC] = \sum_{X,Y,Z}\prob[A|X,Y]\prob[B|Y,Z]\prob[C|Z,X]\prob[X]\prob[Y]\prob[Z]\,. \eq \label{eq:triangle_compatibility} \]
    By contrast, the quantum realization of the Triangle structure are defined with substantially greater freedom, namely
    \[ \prob[ABC] \text{ is a quantum realization of \cref{fig:bell_structure}} \quad\iff\quad\prob[ABC] = \Tr\bs{\netperm^\intercal \rho_{AB}\otimes\rho_{BC}\otimes\rho_{CA} \netperm M_{A}\otimes M_{B} \otimes M_{C}}\,, \eq \label{eq:triangle_Qcompatibility} \]
    where $\rho_{AB}, \rho_{BC}, \rho_{CA}$ are bipartite density matrices, $M_{A}, M_{B}, M_{C}$ are generic measurements sets, and $\netperm$ is a permutation matrix to align the underlying tensor structure of the states and measurements appropriately.

    The Triangle structure serves as an excellent test case for furthering our understanding of quantum nonclassicality in network causal structures. It maintains superficial simplicity (only three observable variables) while introducing many challenging features not found in the study of the Bell structure.
    For example, the spaces of both classical and quantum distributions on the Triangle structure are nonconvex \cite{Fritz_2012,Inflation}, unlike for the Bell structure. The convexity of the Bell structure's distributions is arguably responsible for the wealth of knowledge about it, including its complete characterization of classicality \cite{Brunner_2013}.
    Importantly, \citet{Fritz_2012} explicitly demonstrated the existence of (at least) one incompatible but quantum distribution for the Triangle structure, so it is known to possess quantum nonclassicality. It seems reasonable to assume that quantum nonclassicality in the Triangle structure should be translatable into a computational advantage for certain computational circuits~\cite{Terhal_2002}; novel instances of nonclassicality are expected to correspond to novel information-theoretical quantum advantages. A fundamental limitation of Fritz's proof of nonclassicality, however, is that it does not involve causal compatibility inequalities, and hence does not advance our repertoire of inequality constraints for the Triangle structure.
    Some inequality constraints for the Triangle structure have been derived in previous works. For example, \citet{Steudel_2010} derived an inequality distinguishing the distributions compatible with the Triangle structure from those compatible with structures in which all the observable variables share a common latent ancestor. \citet{Henson_2014} derived a family of entropic inequalities for the Triangle structure, which was the expanded somewhat by \citet{Weilenmann_2016}. Recently, \citet{Inflation} derived a variety of new, especially sensitive, polynomial causal compatibility inequalities for the Triangle structure. In particular, the inequalities of \cite{Inflation} expose a previously unclassified (as assessed by all formerly known constraints) distribution called the $w$-distribution\footnote{Although the $w$-distribution is nonclassical, it is also nonquantum. The nonquantum nature of the $w$-distribution has been demonstrated by Miguel Navascués and Elie Wolfe in private correspondence.} as incompatible with the Triangle structures.
    Remarkably, \emph{none} of the existing causal compatibility inequalities for the Triangle structure were known to admit violation by any quantum distribution. That is to say, it was unknown if any inequality known for the Triangle structure might be useful for distinguishing quantum distributions from their classical counterparts. \Cref{sec:preliminary_research} reports our attempt to utilize these aforementioned inequalities to search for incompatible distributions that are also quantum-accessible. The subsequent failure of these approaches effectively motivates the remainder of this work.
    In summary, the Triangle structure is a desirable case study because it is known to admit nonclassical distributions using quantum resources, but yet no inequality heretofore could separate its classical distributions from its quantum distributions. This failure represents a gap in our understanding of quantum nonclassicality and prompts the discovery of \textit{new} inequalities.

    \section{Fritz Distribution}
    \label{sec:fritz_distribution}
    \begin{figure}
    \begin{nscenter}
        \scalebox{1.0}{\includegraphics{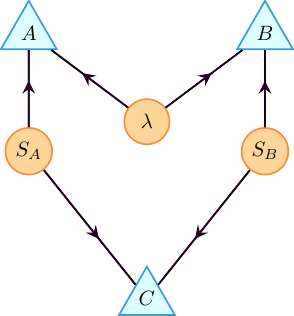}}
        \caption{The Triangle structure reimagined to mimic the Bell structure. The measurement settings $S_{A},S_{B}$ are latent nodes unlike the Bell structure (\cref{fig:bell_structure}).}
        \label{fig:triangle_structure_with_fritz_bell_embedded}
    \end{nscenter}
    \end{figure}

    As first realized by \citet{Fritz_2012}, one may construct quantum distributions incompatible with the Triangle structure by recasting quantum distributions incompatible with the familiar Bell structure into a settings-free tripartite format. To explain, imagine rearranging the Triangle structure into the configuration depicted in \cref{fig:triangle_structure_with_fritz_bell_embedded} so that it closely resembles the Bell structure (\cref{fig:bell_structure}). Evidently, under the correct relabeling, large portions of the Triangle structure resemble the Bell structure. The crucial distinction is that $S_{A}, S_{B}$ are random variable representing the recorded measurement \emph{settings} in the Bell structure whereas those $S_{A}, S_{B}$ are \emph{latent} variables in in the Triangle structure, which get reported as \textit{auxiliary outcomes} for Alice and Bob.

    The analysis of nonclassicality changes, however, when $S_{A}, S_{B}$ are not freely chosen by the observers but rather by a process outside of the individual party's control. Relaxing the assumption of measurement independence opens up a possible loophole, namely the possibility that the auxiliary outcomes $S_{A}, S_{B}$ of Alice and Bob might be manipulated via dependence on their shared latent variable $\lambda$. This loophole is closed by having the third party in the Triangle structure, Charlie, \emph{also} report the latent variables $S_{A}, S_{B}$ as a multivariate outcome. In this manner, the perfect correlation of $C$'s record of $S_{A}, S_{B}$ with the records of $S_{A}$ reported by $A$ and of $S_{B}$ reported by $B$ testifies to the independence of $S_{A}, S_{B}$ from $\lambda$. Consequently, any distribution over $A, B, S_{A}, S_{B}$ that is incompatible with Bell structure is also incompatible with the Triangle structure provided that $C$ is \textit{perfectly correlated} with $S_{A}, S_{B}$~\cite{Fritz_2012}.

\begin{figure}
    \begin{nscenter}
            \includegraphics{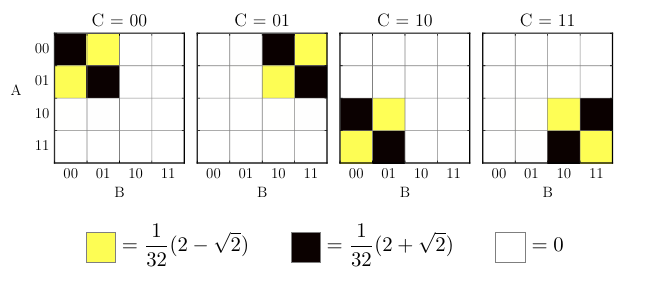}
            \caption{The Fritz distribution visualized using a $4 \times 4 \times 4$ grid. The $4$ outcomes of $A,B,C$ are written in binary as a doublet of bits to illustrate that certain bits act as measurement pseudosettings.}
            \label{fig:fritz_distribution_visualized}
    \end{nscenter}
    \end{figure}

    The exemplifying quantum distribution corresponding to a recasting of a nonclassical Bell structure distribution into the Triangle structure is the \term{Fritz distribution}~\cite{Fritz_2012}. In the Fritz distribution, denoted $\prob[\fritz]$, each of the variables $A,B,C$ is taken to have $4$ possible outcomes $\bc{0,1,2,3}$. Explicitly, $\prob[\fritz]$ can be written as:
    \begin{align*}
    \eq \label{eq:fritz_dist}
    \begin{split}
        \prob[\fritz][000] = \prob[\fritz][110] = \prob[\fritz][021] = \prob[\fritz][131] = \prob[\fritz][202] = \prob[\fritz][312] = \prob[\fritz][233] = \prob[\fritz][323] &= \f{1}{32}\br{2 + \sqrt{2}} \\
        \prob[\fritz][010] = \prob[\fritz][100] = \prob[\fritz][031] = \prob[\fritz][121] = \prob[\fritz][212] = \prob[\fritz][302] = \prob[\fritz][223] = \prob[\fritz][333] &= \f{1}{32}\br{2 - \sqrt{2}}
    \end{split}
    \end{align*}
    In~\cref{eq:fritz_dist}, the notation $\prob[\fritz][abc] = \prob[][A\mapsto a,B\mapsto b,C\mapsto c]$ is used as shorthand. The Fritz distribution is quantum-accessible in the sense that $\prob[\fritz]$ can be implemented using a set of quantum states $\rho_{AB}, \rho_{BC}, \rho_{CA}$ and measurements $M_A, M_B, M_C$ realized on \cref{fig:triangle_structure} using \cref{eq:triangle_Qcompatibility}. When expressing the outcomes $\bc{0,1,2,3}$ as pairs of binary digits $\bc{00, 01, 10, 11}$, it can be seen that the left-hand bits for $A$ and $B$ (respectively denoted $A_l$, $B_l$) are fixed by the outcome of $C$.
    \begin{align*}
    \eq \label{eq:fritz_dist_bit_variant}
    \begin{split}
        \prob[\fritz][00,00,00] = \prob[\fritz][01,01,00] = \prob[\fritz][00,10,01] = \prob[\fritz][01,11,01] &= \f{1}{32}\br{2 + \sqrt{2}} \\
        \prob[\fritz][10,00,10] = \prob[\fritz][11,01,10] = \prob[\fritz][10,11,11] = \prob[\fritz][11,10,11] &= \f{1}{32}\br{2 + \sqrt{2}} \\
        \prob[\fritz][00,01,00] = \prob[\fritz][01,00,00] = \prob[\fritz][00,11,01] = \prob[\fritz][01,10,01] &= \f{1}{32}\br{2 - \sqrt{2}} \\
        \prob[\fritz][10,01,10] = \prob[\fritz][11,00,10] = \prob[\fritz][10,10,11] = \prob[\fritz][11,11,11] &= \f{1}{32}\br{2 - \sqrt{2}}
    \end{split}
    \end{align*}
    In~\cref{eq:fritz_dist_bit_variant}, the notation $\prob[\fritz][a_{l}a_{r},b_{l}b_{r},c_{l}c_{r}] = \prob[][A_{l}\mapsto a_{l},A_{r}\mapsto a_{r},B_{l}\mapsto b_{l},B_{r}\mapsto b_{r},C_{l}\mapsto c_{l},C_{r}\mapsto c_{r}]$ is used as shorthand. This observation can be difficult to verify using~\cref{eq:fritz_dist_bit_variant}, but becomes easier after organizing the possible outcomes into a $4 \times 4 \times 4$ grid as depicted in \cref{fig:fritz_distribution_visualized}. From this diagram, it can be seen that each of $C$'s outcomes restricts the possible outcomes for $A,B$ into a $2 \times 2$ block. Effectively, $C$'s bits are perfectly correlated with the left-hand bits of $A,B$; $C_l = A_l$ and $C_r = B_l$.
    Therefore, pursuant to the embedding of~\cref{fig:triangle_structure_with_fritz_bell_embedded}, the left-hand bits $A_l$ and $B_l$ emulate the measurement settings $S_{A}$ and $S_{B}$, whereas the right-hand bits $A_r$ and $B_r$ emulate the outcomes which would be obtained by $A$ and $B$ back in the Bell structure (\cref{fig:bell_structure}). Provided that $C$ is perfectly correlated with $A_l$ and $B_l$, any Bell inequality for the Bell structure defined over $A, B, S_{A}, S_{B}$ can be directly converted to an inequality for the Triangle structure by performing the surjective relabeling:
    \begin{align*}
        A_r\leftarrow A\,,\quad
        B_r\leftarrow B\,,\quad
        A_l\leftarrow S_{A}\,,\quad
        B_l\leftarrow S_{B}\,,\quad
        C_l\leftarrow S_{A}\,,\quad
        C_r\leftarrow S_{B}\,.
        \eq \label{eq:fritz_relabelling}
    \end{align*}
    As an example, the famous CHSH inequality (\cref{eq:CHSH_ineq_orig})~\cite{CHSH_Original} is translated into a constraint on the correlation between the right-hand bits of $A$ and $B$ under~\cref{eq:fritz_relabelling},
    \[ C = C_{l}C_{r} = A_{l}B_{l} \Longrightarrow \ba{A_r B_r|C = 00} + \ba{A_r B_r|C = 01} + \ba{A_r B_r|C = 10} - \ba{A_r B_r|C = 11} \leq 2, \eq \label[ineq]{eq:CHSH_intriangle} \]
    where the correlation $\ba{A_r B_r}$ between the right-hand bits of $A$ and $B$ is defined as,
    \[ \ba{A_r B_r} = \prob[A_r B_r][00] + \prob[A_r B_r][11] - \prob[A_r B_r][01] - \prob[A_r B_r][10]. \eq \label{eq:right_bit_correlations} \]
    Therefore, every compatible distribution $\prob[ABC]$ \emph{for which $C$ is perfectly correlated with $A_l,B_l$} must satisfy \cref{eq:CHSH_intriangle}. Substituting $\prob[\fritz]$ into \cref{eq:CHSH_intriangle} yields the traditional maximal quantum violation~\cite{Cirelson_1980}: $3 \br{1/\sqrt{2}} - \br{-1/\sqrt{2}} = 2\sqrt{2} \not \leq 2$.

    It is important to understand the domain in which Fritz's proof of incompatibility is valid; its proof relies on the perfect correlation between $C$'s outcomes and the measurement pseudosettings (left-hand bits) of $A$ and $B$. For example, if one combines \cref{eq:fritz_dist} with slight uniform noise, what can be said with confidence regarding if the resulting modified distribution is classical or not? At what point does the resulting distribution transition from incompatibility to compatibility? This question is partially answered in \cref{sec:noise}.

    Plainly, $\prob[\fritz]$ is a valid but \textit{manufactured} example. The phenomenology associated with Bell nonlocality or Bell incompatibility are well understood; examining these distributions embedded in the Triangle structure offers no additional perspective onto the types of nonclassical resources made accessible by quantum mechanics. The goal, therefore, is to find incompatible quantum distributions that are qualitatively different than those previously considered for the Bell structure \cite{Gisin_2017}. Recognizing this, \citet{Fritz_2012} presented the following problem~\cite[Problem 2.17]{Fritz_2012}:

    \term{Fritz's Problem}: Find an example of nonclassical quantum correlations in the Triangle structure together with a proof of its nonclassicality which does not hinge on Bell’s theorem.

    Fritz's problem is concerned with how to find and recognize nonclassical quantum distributions specifically for the Triangle structure. The original proof of nonclassicality essentially recycled Bell's theorem, and was limited by the requirement of perfect correlations~\cite{Fritz_2012}. Fritz's problem, as originally stated, does not require that the \emph{type} of nonclassicality be novel to the Triangle structure, rather, only that the proof should avoid Bell's theorem. \Cref{sec:preliminary_research} delineates our initial, failed attempts at resolving Fritz's problem; \Cref{sec:found_inequalities} reports our eventual success, via the discovery of different causal compatibility inequalities.

    Though not explicit, we read in the spirit of Fritz's problem a desideratum for the discovery of a truly different form of nonclassicality for the Triangle structure. Such a discovery would presumably lead to an understanding of different advantages of quantum resources in network structures; this related problem has attracted attention and conjecture \cite{Gisin_2017}, but remains open. We attempt to make progress on this problem by leveraging the causal compatibility inequalities derived herein, but \Cref{sec:optimizations} delineates that is effort was plagued by instabilities in our numerical optimization which we have not yet overcome. Consequently, \Cref{sec:fritz_problem_revisit} discusses potential avenues for rigorously reformulating Fritz's problem in order to best capture this desire for novel quantum nonclassicality.

    \section{Preliminary Research}
    \label{sec:preliminary_research}

    As a preliminary search for quantum incompatibility in the Triangle structure, we performed numerical optimizations (in search of violation) against the previously published compatibility inequalities of~\citet{Inflation}, as well as against the entropic inequalities of \citet{Henson_2014}. For historical context, the entropic inequalities of \cite{Henson_2014} have already been independently investigated for quantum incompatibility by \citet{Weilenmann_2016} using a variety of computational methods. Unfortunately, these methods failed to identify quantum-accessible distributions capable of violating any of the entropic inequalities considered by \cite{Weilenmann_2016}. Additionally, the inequalities presented in~\cite{Inflation} have not been previously investigated numerically.

    Explicitly, we parameterized the subset of quantum distributions accessible by bipartite qubit density matrices and two-outcome POVM measurements\footnote{In \cref{sec:optimizations}, we discuss methods for conducting similar parameterizations of quantum states and measurements.}. This preliminary investigation did not yield interesting solutions, as none of the Triangle structure inequalities in~\cite{Inflation} or~\cite{Henson_2014} were violated\footnote{We also directly checked all two-outcome coarse grainings of the Fritz distribution against the inequalities in~\cite{Inflation,Henson_2014}, with no violation.}. Unfortunately, these early results are inconclusive for two reasons. First, an exhaustive search would need to consider the possibility of larger Hilbert spaces for the shared quantum states. Second, it is known that the inequalities~\cite{Inflation} are incomplete; there exists other constraints on two outcome distributions for the Triangle structure yet to be discovered.

    Nonetheless, having failed to observe a quantum-classical gap in the Triangle structure for \textit{binary-outcome} measurements, a continued search for nonclassical distributions in the Triangle structure must expand the gamut of inequalities to optimize against to include inequalities referencing strictly more than two outcomes.

    \section{Triangle Structure Inequalities}
    \label{sec:found_inequalities}

    \Cref{sec:causal_compatibility} introduced the notion of causal compatibility inequalities, and \cref{sec:fritz_distribution} discussed the Fritz distribution ($\prob[\fritz]$) together with the initial inequality-free proof of its incompatibility with the Triangle structure. Heretofore, there were no known causal compatibility inequalities for the Triangle structure that were capable of witnessing the nonclassicality of any quantum distributions~\cite{Inflation}. By leveraging the incompatibility of the Fritz distribution~\cite{Fritz_2012} and tools provided by the Inflation Technique~\cite{Inflation}, we have obtained numerous causal compatibility inequalities for the Triangle structure that are violated by the Fritz distribution. A representative trio of these $\prob[\fritz]$-incompatibility-witnessing inequalities are presented here: $I\tsb{WagonWheel}$ per \cref{eq:ww_ineq}, notable for its simplicity; $I\tsb{Web}$ per \cref{eq:web_inequality}, which best witnesses the nonclassicality of $\prob[\fritz]$ in the presence of noise; and $I\tsb{SymmetricWeb}$ per \cref{eq:symmetric_web_inequality}, which is symmetric with respect to all permutations of the three parties.

    Readers which are unfamiliar with the Inflation Technique~\cite{Inflation}, and wish to understand in detail how the following inequalities are derived, are recommended to consult \cref{sec:marginal_satisfiability,sec:inflation_technique_summary,sec:deriving_inequalities} wherein they will find a succinct yet sufficient presentation of the requisites needed for this paper. To briefly summarize, \cref{sec:inflation_technique_summary} reviews the basics of the Inflation Technique and formally defines the notion of an \textit{inflation} of a causal structure. \Cref{sec:deriving_inequalities} demonstrates how to use the Inflation Technique and a given probability distribution, such as the Fritz distribution, to cast the causal compatibility problem as a particular kind of linear program known as a marginal problem, defined in \cref{sec:marginal_satisfiability}.

    The Inflation Technique is known to completely solve the causal compatibility problem through increasing orders of inflations~\cite{Navascues_2017}. Consequently, the derivation of \cref{eq:ww_ineq,eq:web_inequality,eq:symmetric_web_inequality} is guaranteed by~\cite{Navascues_2017} for sufficiently large inflations. Nonetheless, the following inequalities were obtained using the relatively small inflations found in \cref{fig:inflations}\footnote{Note that the adjectives \textit{large} and \textit{small} used to describe an inflation only become well defined in the context of the hierarchy proposed in~\cite{Navascues_2017}. For example, the Web Inflation in \cref{fig:the_web_inflation} is second in the hierarchy of~\cite{Navascues_2017} whereas the Wagon-Wheel Inflation in \cref{fig:wagon_wheel_inflation} is somewhere between the first and second order.}. This efficiency is not universally guaranteed; for comparison, we remark that unlike the Fritz distribution, the conjectured incompatibility of the four-outcome distribution proposed by \cite{Gisin_2017} is \textit{not} confirmed by Inflation Technique at the same level.

    We emphasize that each of these causal compatibility inequalities independently provides a positive resolution of Fritz's problem. The inequalities are derived without making use of Bell's theorem; rather, they follow from the Inflation Technique's broader perspective on nonclassicality as a special case of causal inference. Afterwards, Section~\ref{sec:fritz_problem_revisit} returns to the topic of Fritz's problem and the status of our resolution.

    \subsection{The WagonWheel Inequality}
    The first causal compatibility inequality chosen for presentation is reported below:\footnote{\label{foot:error}In earlier versions of this paper, including the version published in Physical Review A (found here: \url{https://journals.aps.org/pra/abstract/10.1103/PhysRevA.98.022113}) and the initial arXiv submission, \Cref{eq:ww_ineq} was incorrectly presented. A detailed errata can be found here: \url{https://journals.aps.org/pra/abstract/10.1103/PhysRevA.102.029901}.} 
    
    \begin{equation*}
    \begin{gathered}
        \eq\label[ineq]{eq:ww_ineq}
        I\tsb{WagonWheel} : \\
        -P_{A_{l}B_{l}}(10)
        +P_{A_{l}B_{l}C_{l}C_{r}}(1010)
        -P_{A_{l}B_{l}}(01) P_{C_{l}C_{r}}(10)
        -P_{C_{l}C_{r}}(00) P_{C_{l}C_{r}}(11) \\
        -P_{C_{l}C_{r}}(01) P_{A_{l}A_{r}B_{l}B_{r}C_{l}C_{r}}(100110)
        -P_{C_{l}C_{r}}(01) P_{A_{l}A_{r}B_{l}B_{r}C_{l}C_{r}}(110010) \\
        +P_{C_{l}C_{r}}(00) P_{A_{l}A_{r}B_{l}B_{r}C_{l}C_{r}}(101111)
        +P_{C_{l}C_{r}}(00) P_{A_{l}A_{r}B_{l}B_{r}C_{l}C_{r}}(111011) \\
        +P_{C_{l}C_{r}}(10) P_{A_{l}A_{r}B_{l}B_{r}C_{l}C_{r}}(001001)
        +P_{C_{l}C_{r}}(10) P_{A_{l}A_{r}B_{l}B_{r}C_{l}C_{r}}(011101) \\
        +P_{C_{l}C_{r}}(11) P_{A_{l}A_{r}B_{l}B_{r}C_{l}C_{r}}(000000)
        +P_{C_{l}C_{r}}(11) P_{A_{l}A_{r}B_{l}B_{r}C_{l}C_{r}}(010100) \\
        \leq 0
    \end{gathered}
    \end{equation*}

    \Cref{eq:ww_ineq} is termed the Wagon-Wheel inequality and denoted $I\tsb{WagonWheel}$ after the (identically named) inflated structure of \cref{fig:wagon_wheel_inflation} used to derive it. To reiterate, a summary of the methods used to derive \cref{eq:ww_ineq} can be found in \cref{sec:marginal_satisfiability,sec:inflation_technique_summary,sec:deriving_inequalities}. Moreover, note that \Cref{eq:ww_ineq} is reported using the same two-bit notation discussed in~\cref{sec:fritz_distribution} such that $\prob[A_{l}A_{r}B_{l}B_{r}C_{l}C_{r}][a_{l}a_{r}b_{l}b_{r}c_{l}c_{r}] = \prob[ABC][abc]$.

    Aside from increasing outcome cardinality, we are also forced to consider larger inflations than those analyzed in~\cite{Inflation}. This is because we found the smaller inflations considered there, such as the Spiral inflation depicted in \cref{fig:spiral_inflation}, were simply \textit{unable} to witness the incompatibility of the Fritz distribution, even when analyzed explicitly using four possible outcomes for every observable variable.

    By construction, every distribution $\prob[ABC]$ that is compatible with the Triangle structure per \cref{eq:triangle_compatibility} must satisfy $I\tsb{WagonWheel}$. On the other hand, the Fritz distribution violates $I\tsb{WagonWheel}$ with violation $\f{1}{16}(\sqrt{2} - 1) \not \leq 0$\footnote{As a result of the error mentioned in Footnote~\ref{foot:error}, earlier versions of this paper erroneously reported a numerical violation of $\f{1}{16} \not \leq 0$. }. Consequently we affirm that the Fritz distribution is incompatible with the Triangle structure, a fact previously only demonstrated without inequalities.

    \subsection{The Web Inequality}
    \label{sec:noise}
    \label{sec:violations_noise}
    We next present \cref{eq:web_inequality}, another causal compatibility inequality for the Triangle structure that is violated by the Fritz distribution. The Web inflation shown in \cref{fig:the_web_inflation} was used to produce the eponymous $I\tsb{Web}$ per \cref{eq:web_inequality}.  The Web inflation is considerably larger than the Wagon-Wheel inflation; it is computationally more demanding to work with, albeit capable of yielding strictly stronger inequalities. For brevity, we employ the shorthand $\prob\br{abc}$ in lieu of $\prob[ABC]\br{abc}$ in presenting \cref{eq:web_inequality}.

    \begin{gather}
    \label[ineq]{eq:web_inequality}
        I\tsb{Web} : \\\nonumber
        \includegraphics[clip, trim=0cm 0cm 0cm 0.5cm, width=\textwidth]{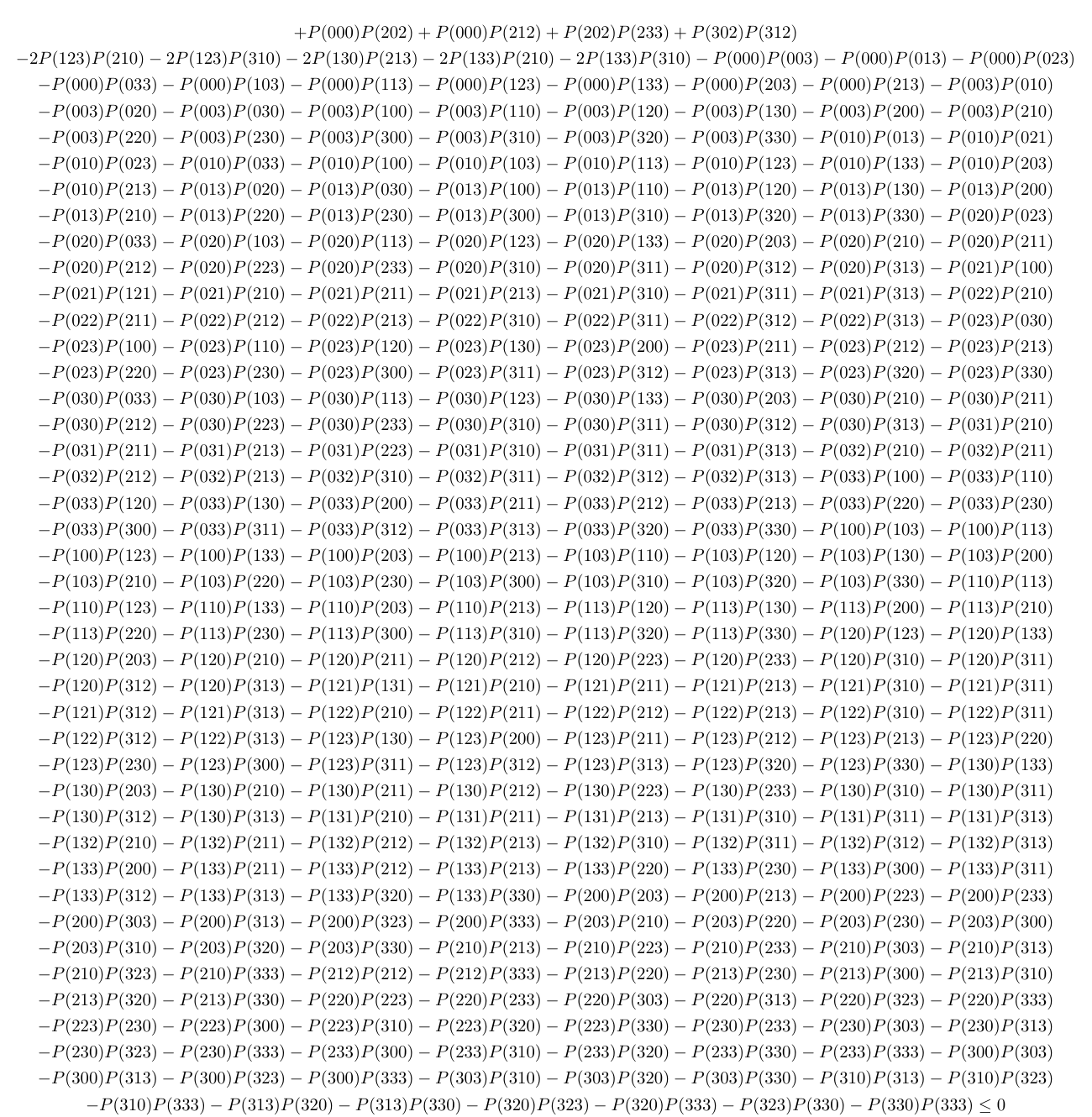}
    \end{gather}

    As mentioned in \cref{sec:fritz_distribution}, this original proof of the nonclassicality of $\prob[\fritz]$ required an idealistic condition to hold: perfect correlations of $C$ with $A_l, B_l$. It is impossible to use Fritz's original argument to confirm nonclassicality from an experimental point of view, because every laboratory-achievable distribution is subject to some amount of \text{noise}. One can minimize noise. e.g. by developing high-accuracy measurement channels, but \emph{perfect} correlations are unattainable. Causal compatibility inequalities such as  $I\tsb{Web}$ permit there to be noise within a set of observations before the ability to certify nonclassicality breaks down.

    Here we quantify how much statistical noise can be added to the Fritz distribution $\prob[\fritz]$ before $I\tsb{Web}$ fails to witness incompatibility; we define the \term{$\vep$-noisy Fritz distribution} as
    \begin{align}
        \begin{split}
            \s{N}_{\vep} = \br{1 - \vep} \prob[\fritz] + \vep \s{U}_{ABC} \quad \text{where} \quad \s{U}_{ABC}\br{abc} \equiv \f{1}{64} \quad \forall a,b,c\in \bc{0,1,2,3}\,. \\
        \end{split}
    \end{align}

\begin{figure}
    \begin{nscenter}
            \includegraphics{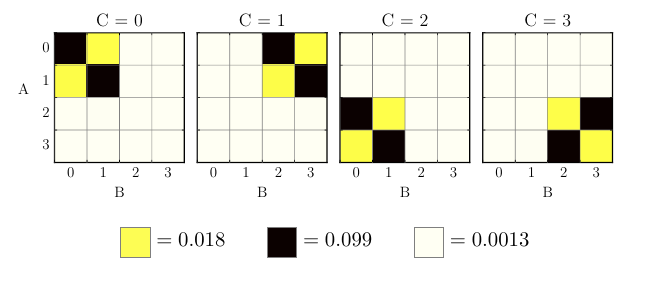}
            \caption{$\s{N}_{0.085}$: A noisy yet still nonclassical variant of the Fritz distribution.}
            \label{fig:noisy_fritz}
    \end{nscenter}
    \end{figure}

    As $\vep$ varies from $0$ to $1$, noise is added to the Fritz distribution and $\s{N}_{\vep}$ transitions from an incompatible distribution $\s{N}_{0} = \prob[\fritz]$ to a compatible distribution $\s{N}_{1} = \s{U}_{ABC}$. We find that $I\tsb{Web}$ demonstrates that the Fritz distribution remains incompatible with the Triangle structure up to a noise parameter of $\vep \simeq 0.085$; the associated distribution $\s N_{0.085}$ is plotted in \cref{fig:noisy_fritz}. Of course, there remains the possibility that another inequality will be able to withstand a larger degree of noise than $I\tsb{Web}$; an exhaustive search has not been conducted.

    \subsection{The Symmetric Web Inequality}
    In \cref{sec:optimizations} we will consider numerically optimizing our inequalities over quantum strategies, seeking violations, towards the desired end goal of discovering different forms of nonclassicality in the Triangle structure. Of course, if some inequality achieves its optimal quantum violation on a distribution qualitatively similar to $\prob[\fritz]$, then such an inequality is unlikely to lead us to discover nonclassicality of a type different than $\prob[\fritz]$. Indeed, we will show that the previous pair of $I\tsb{WagonWheel}$ and $I\tsb{Web}$, which were both specifically curated to demonstrate the incompatibility of $\prob[\fritz]$, apparently achieve maximum quantum violation on distributions very similar to $\prob[\fritz]$ itself.

    In order for the numerical optimization to avoid distributions similar to $\prob[\fritz]$ we require an objective function which is \emph{not} tailor designed to witness $\prob[\fritz]$. Finding such an inequality proved extremely challenging; while we could generate thousands of inequalities using the Inflation Technique~\cite{Inflation}, it appeared than only a vanishingly small fraction of those inequalities admitted quantum violation whatsoever. We therefore modified the Inflation Technique to give us an inequality which both witnesses the nonclassicality of $\prob[\fritz]$ but which is also symmetric with respect to any permutation of the variables $A, B, C$, resulting in \cref{eq:symmetric_web_inequality}. The modification of the Inflation Techniques which forces symmetry is explained in \cref{sec:symmetric_inequalities}. As $\prob[\fritz]$ is strongly asymmetric due to the special role of $C$, we can hope that a symmetric inequality (even one which does witness $\prob[\fritz]$) might achieve its \emph{optimal} quantum violation on a distribution qualitative distinct from $\prob[\fritz]$. The details of our numerical finding are discussed in \cref{sec:optimizations}.

    \begin{gather}
        \label[ineq]{eq:symmetric_web_inequality}
        I\tsb{SymmetricWeb} : \\\nonumber
        \includegraphics[clip, trim=0cm 0cm 0cm 0.5cm, width=\textwidth]{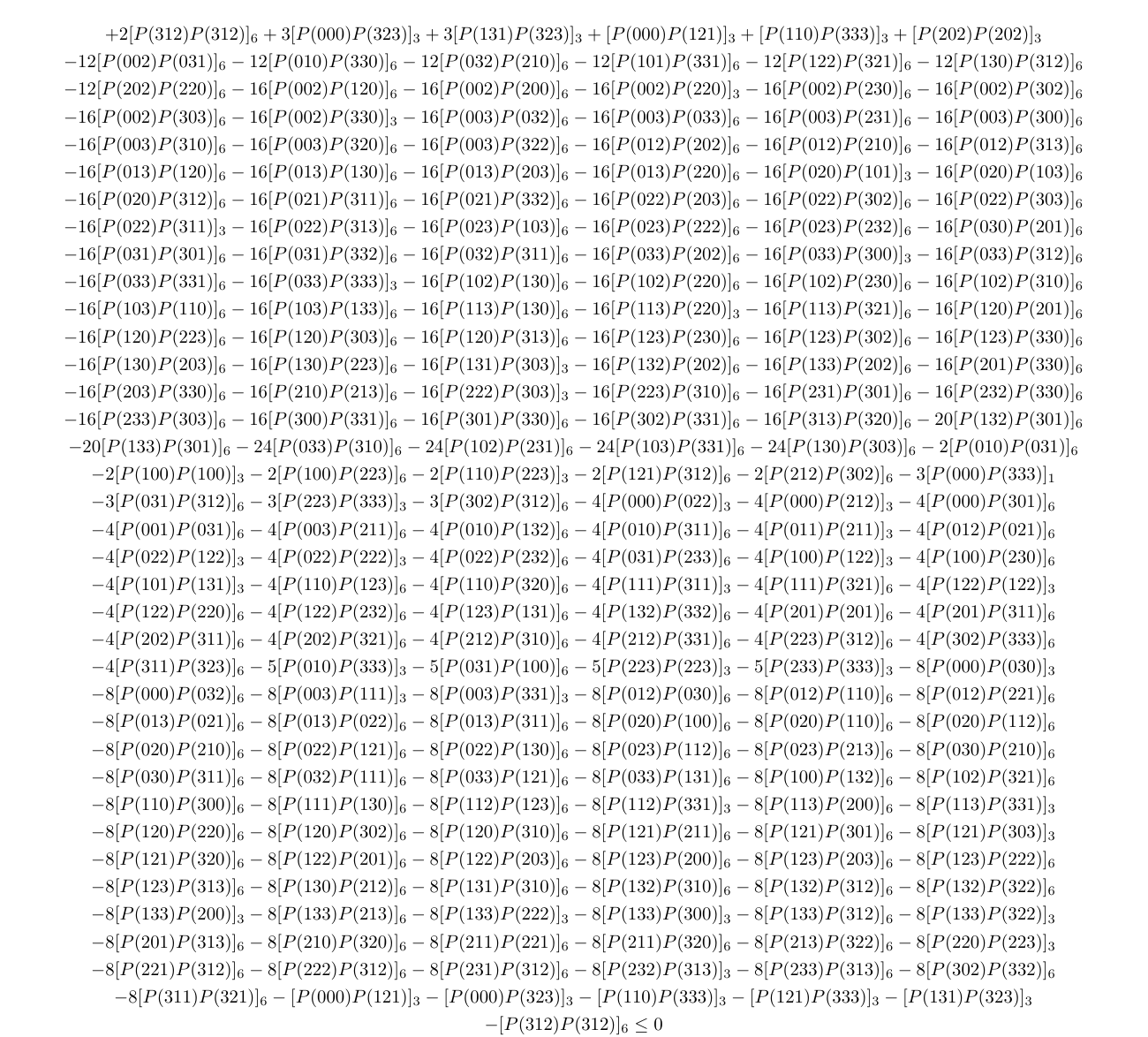}
    \end{gather}

    Note that $\prob\br{abc}$ is shorthand for $\prob[ABC]\br{abc}$ and $\bs{P(113)P(330)}_3$ is shorthand for a sum the over permutations of $A,B,C$, e.g. $\bs{P(113)P(330)}_3 \defined P(113)P(330) + P(131)P(303) + P(311)P(033)$.

    \section{Numerical Optimization}\label{sec:optimizations}

    In order to find quantum distributions that are more nonclassical than the Fritz distribution, we performed numerical optimizations against each inequality $I$ by parameterizing the space of quantum-accessible probability distributions that can be realized on the Triangle structure (\cref{fig:triangle_structure}) and thus expressed in the form of \cref{eq:triangle_Qcompatibility}. In order to parameterize all such distributions, we elect to parameterize the states and measurements separately. In order to qualify the scope of \cref{eq:triangle_Qcompatibility} and associated computational complexity of the parameterization, there are a two restrictions that are made with justification. Motivated by the fact that the Fritz distribution (\cref{sec:fritz_distribution}) only requires qubit states, the states $\rho$ are taken to be bipartite \textit{qubit} states which are more computationally feasible compared to $n$-dimensional states whereby the joint density matrix $\rho_{AB}\otimes\rho_{BC}\otimes\rho_{CA}$ becomes an $n^6 \times n^6$ matrix. Additionally, we restrict our focus to projective-valued measures (PVMs) instead of projective-operator valued measures (POVMs) for three reasons. First,~\citet{Fritz_2012} demonstrates via the Fritz distribution that PVMs are sufficient for generating incompatible quantum distributions in the Triangle structure. Second, although generating $k$-outcome POVM measurements is possible using rejection sampling techniques~\cite{Petz_2015}, a valid, unbiased parameterization was not found for $k > 2$. Finally, PVMs provide considerable computational advantage over POVMs as they permit \cref{eq:triangle_Qcompatibility} to be rewritten as \cref{eq:quantum_model_triangle_pvms}.
    \[ \prob[ABC]\br{abc} = \bramidket{m_{A,a}m_{B,b}m_{C,c}}{\netperm^\intercal \rho_{AB}\otimes\rho_{BC}\otimes\rho_{CA} \netperm}{m_{A,a}m_{B,b}m_{C,c}} \eq \label{eq:quantum_model_triangle_pvms}\]

    Although there are numerous techniques that can used when parameterizing quantum states and measurements~\cite{Petz_2015, Hedemann_2013,Fujii_2005,James_2001,Grasmair_2014,Neilsen_Chaung_2011}, a single technique by \citet{Spengler_2010_Unitary} was found to be most computationally suitable for our purposes. \Citet{Spengler_2010_Unitary} demonstrated that all $d\times d$ unitary matrices $U$ can be parameterized without degeneracy as follows:
    \[ U = \bs{\prod_{m=1}^{d-1} \br{\prod_{n=m+1}^{d} \exp\br{i P_n \lambda_{n,m}}\exp\br{i \si_{m,n} \lambda_{m,n}}}} \times \bs{\prod_{l=1}^{d} \exp\br{iP_l \lambda_{l,l}}}  \eq \label{eq:spengler_unitary} \]
    Where the real valued parameters $\la = \bc{\la_{n,m} \mid n,m \in 1, \ldots, d}$ have periodicities $\la_{m,n} \in \bs{0, \f{\pi}{2}}$ for $m < n$ and $\la_{m,n} \in \bs{0, 2 \pi}$ for $m \geq n$. Moreover, $P_l$ are one-dimensional projective operators $P_l = \ket{l}\bra{l}$ and the $\si_{m,n}$ are generalized antisymmetric $\si$-matrices $\sigma_{m,n} = -i \ket{m}\bra{n} +i \ket{n}\bra{m}$ where $1 \leq m < n \leq d$. This parameterization has the useful feature that each of the real-valued parameters $\la_{n,m}$ a direct and intuitive physical affect on each element of a computational basis $\bc{\ket{1}, \ldots, \ket{d}}$. Explicitly, $\exp\br{i \si_{m,n} \lambda_{m,n}}$ applies a rotation to the subspace spanned by $\ket{m}$ and $\ket{n}$ for $m < n$. Analogously, $\exp\br{i P_n \lambda_{n,m}}$ generates the relative phase between $\ket{m}$ and $\ket{n}$ for $m > n$ and $\exp\br{iP_l \lambda_{l,l}}$ fixes the global phase of $\ket{l}$. Finally, although not explicitly mentioned in~\cite{Spengler_2010_Unitary}, it is possible to remove the reliance on the computationally expensive matrix exponential operations~\cite{Moler_2003} in \cref{eq:spengler_unitary} and replace them with elementary trigonometric functions in the parameters $\la_{m,n}$.

    By parameterizing unitary matrices, it becomes possible to parameterize $d$-dimensional density matrices and $d$-element PVMs by recognizing that any orthonormal basis $\bc{\ket{\psi_j}}$ (where $1 \leq j \leq d$) can be transformed into the computational basis $\bc{\ket{j}}$ by a unitary transformation $U$, i.e. $U \ket{\psi_j} = \ket{j}$. First consider a $d$-element PVM $M = \bc{\ket{m_j}\bra{m_j} \mid 1 \leq j \leq d}$. Since $\bc{\ket{m_j}}$ forms an orthonormal basis, one can parameterize $M$ by writing $M = \bc{U^{\dagger}\ket{j}\bra{j}U \mid 1 \leq j \leq d}$ and parameterizing $U$ using \cref{eq:spengler_unitary}. This method was inspired by the measurement seeding method for iterative optimization used by \citet{Pal_2010}. Analogously, this argument can be extended to full-rank $d$-dimensional density matrices $\rho$ by performing a spectral decomposition $\rho = \sum_{j=1}^{d} p_j \ket{p_j} \bra{p_j}$ into eigenvalues $\bc{p_j}$ and eigenstates $\bc{\ket{p_j}}$. Since $\Tr\br{\rho} = \sum_{j = 1}^{d} p_j = 1$ the eigenvalues of $\rho$ are parameterized without degeneracy using a tuple of $d-1$ real-valued parameters with periodicity $\bs{0, 2 \pi}$ using hyper-spherical coordinates~\cite{Hedemann_2013, Spengler_2010_Unitary}. Additionally, since $\rho$ is Hermitian, the eigenstates $\bc{\ket{p_j}}$ form an orthonormal basis and therefore the eigenstates are analogously parameterized using \cref{eq:spengler_unitary}: $\rho = \sum_{j=1}^{d} p_j U^{\dagger} \ket{j} \bra{j} U$. For our purposes, we have set $d = 4$ and fixed $\la_{l, l} = 0$ for $1 \leq l \leq d$ in \cref{eq:spengler_unitary} because the global phase contributions are irrelevant for \cref{eq:quantum_model_triangle_pvms}.

    Since the inequalities we seek to optimally violate are polynomial, and since the space of quantum-accessible distributions is nonconvex, we employ a number of optimization methods consecutively in an attempt to avoid pitfalls associated identifying local minima and ill-conditioned convergence. Specifically, the Broyden-Fletcher-Goldfarb-Shanno method~\cite[p.142]{Nocedal_2000} and the Nelder-Mead simplex method~\cite[p.238]{Nocedal_2000} were used along with a method called Basin Hopping~\cite{Wales_1997}, which is a hybrid between simulated annealing and gradient-descent-based methods.

    Nevertheless, we strongly caution against misinterpreting the numerical optima present in \cref{sec:optima} as if our findings represent genuine global maxima of violation. Evidence for the unreliability of our numerical methods is the following: When supplied with randomly sampled initial parameters, \emph{all} optimization methods consistently converged to \emph{saturating} the target inequalities, instead of violating it! This, even though we \emph{know} that all the inequalities we consider \emph{do} admit quantum violation, namely by  $\prob[\fritz]$. To achieve inequality violation, we found ourselves forced to initialize the numerical optimizer with parameters that were nearby (in parameter space) to parameters which generate the Fritz distribution $\prob[\fritz]$. Upon doing so, it was observed that the numerical methods converged invariably to parameters which generated distributions similar to $\prob[\fritz]$ (see~\cref{fig:maximum_violation_I_1} for instance); consequently making interpreting the results somewhat murky. Could it be that the inequality's global maximum is in fact not far from $\prob[\fritz]$? Or is this a limitation of the ill-conditioned nature of the optimization? The phenomena is likely due to a combination of both effects.

    \subsection{Numerical Optimization Results}\label{sec:optima}

    \begin{figure}
    \begin{nscenter}
        \includegraphics{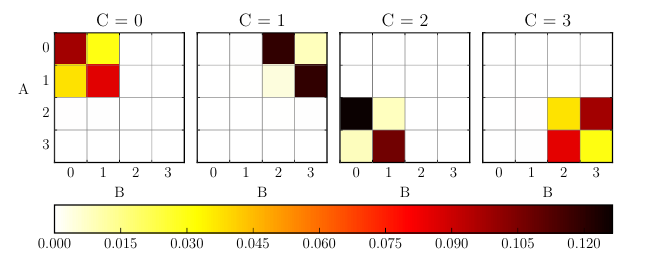}
        \caption{A quantum probability distribution of the Triangle structure that maximizes violation of $I\tsb{Web}$. Notice that this distribution has precisely the same possibilistic structure as the Fritz distribution, as might be expected, since $I\tsb{Web}$ was specifically generated to witness the nonclassicality of $\prob[\fritz]$.}
        \label{fig:maximum_violation_I_1}
    \end{nscenter}
    \end{figure}
    \begin{figure}
    \begin{nscenter}
        \includegraphics{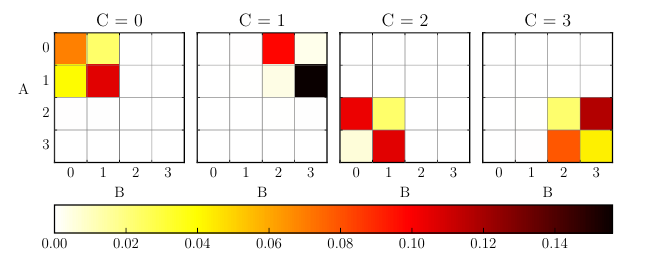}
        \caption{A quantum probability distribution of the Triangle structure that maximizes violation of $I\tsb{SymmetricWeb}$. Notice that this distribution has precisely the same (asymmetric!) possibilistic structure as the Fritz distribution, even though $I\tsb{SymmetricWeb}$ itself is symmetric with respect to permutations of the variables $A, B, C$. }
        \label{fig:maximum_violation_I_3}
    \end{nscenter}
    \end{figure}

    Our best numerical optimization of $I\tsb{WagonWheel}$ was found to be the Fritz distribution itself, visualized in \cref{fig:fritz_distribution_visualized}. $\prob[\fritz]$ is therefore either a local minimum of the parameter space or is in fact \textit{the} maximally violating distribution of $I\tsb{WagonWheel}$; it remains unclear to us if this behavior is related to the global optimality of the Fritz distribution, or if it's an artifact of the methods used to derive $I\tsb{WagonWheel}$, though we suspect the former.

    Our best numerical optimization of $I\tsb{Web}$ is visualized in \cref{fig:maximum_violation_I_1}. Almost immediately, it is evident that \cref{fig:maximum_violation_I_1} closely resembles the Fritz distribution; in fact \cref{fig:maximum_violation_I_1} and \cref{fig:fritz_distribution_visualized} share their \term{possibilistic structure}. A distribution's possibilistic structure is the subset of events which it assigns a nonzero probability. That \cref{fig:maximum_violation_I_1} and \cref{fig:fritz_distribution_visualized} have the same possibilistic structure is not entirely unexpected;  As was mentioned in \cref{sec:found_inequalities}, as $I\tsb{Web}$ was derived specifically to prove the incompatibility of the Fritz distribution with the Triangle structure. Consequently, it seems that $I\tsb{Web}$ best witnesses incompatibility of those distributions that closely resemble the $\prob[\fritz]$. It could be, however, that the sharing of possibilistic structure is an artifact of unreliable convergence, such that the true optimally violating $I\tsb{Web}$ quantum distribution might have a different structure.

    It is perhaps curious, however, that violation achieved by the distribution in \cref{fig:maximum_violation_I_1} is not accessible when the bipartite states in \cref{eq:triangle_Qcompatibility} are restricted to maximally entangled states. This finding of more violation with less entanglement resembles a feature of quantum mechanics originally presented by~\citet{Methot_2006} demonstrating that entanglement and nonclassicality are different resources.

    Our best numerical optimization of $I\tsb{SymmetricWeb}$ is visualized in \cref{fig:maximum_violation_I_3}. Counterintuitively, the distribution in \cref{fig:maximum_violation_I_3}, which achieves a greater violation of $I\tsb{SymmetricWeb}$ than $\prob[\fritz]$, shares its \textit{possibilistic} structure with the Fritz distribution! Additionally, since $I\tsb{SymmetricWeb}$ is symmetric with respect to the permutation of parties, one might expect that the maximum violating distribution be symmetric as well. To the contrary, our numerical optimization failed to find \emph{any} symmetric quantum distributions capable of violating $I\tsb{SymmetricWeb}$.

    \section{Revisiting Fritz's Problem}
    \label{sec:fritz_problem_revisit}

    In~\cref{sec:fritz_distribution}, we defined the Fritz distribution and summarized the inequality-free proof of its incompatibility with the Triangle structure due to~\cite{Fritz_2012}. In addition, \cref{sec:fritz_distribution} discusses Fritz's Problem as a quest to find quantum distributions incompatible with the Triangle structure that are also qualitatively distinct from those distributions which are incompatible with the Bell structure. In light of the perhaps unsatisfactory results of~\cref{sec:optimizations}, the purpose of this section is to revisit Fritz's problem and to attempt to rigorously formulate criterion for quantum nonclassicality that is genuine to the Triangle structure.

    When specifically concerned with entanglement resources, Bell structure nonclassicality exploits the entanglement of a \textit{single} bipartite quantum state. Unlike the Bell structure in~\Cref{fig:bell_structure}, the observable nodes of the Triangle structure in~\Cref{fig:triangle_structure} have access to potentially \textit{three} different bipartite quantum states. As noted in~\cite{Fritz_2012}, the Fritz distribution can be implemented using only a \textit{single} entangled state.\footnote{Of course there exists implementations that make use of three entangled states but at \textit{minimum}, one is necessary.} Therefore, it becomes reasonable to propose that any quantum nonclassicality in the Triangle structure that \textit{requires} entanglement in \textit{at least two} shared states constitutes nonclassicality distinct from Bell nonclassicality. At first, one might suspect that finding such distributions would affirmatively demonstrate a form of entanglement resource unique to the Triangle structure. Unfortunately, this is not the case. By increasing the cardinality of each variable $A, B, C$ to $4^{3} = 64$, it is possible to generate a distribution that is nonclassical yet quantum and requires \textit{all three} of the shared resources to be entangled. This can be accomplished by superimposing three copies of statistically independent Fritz distributions to the Triangle structure, each of which utilizes a distinct party to announce the corresponding measurement pseudosettings. Such a distribution would require entanglement in all three quantum states; under this construction, the removal of any entangled resource would render the distribution quantum-inaccessible\footnote{This observation was original provided by Miguel Navascués.}. Consequently, demanding the necessity of entanglement in every shared state is insufficient for finding novel nonclassicality.

    In consideration of this particular construction, perhaps the correct assessment for novel nonclassicality must incorporate a restriction on the the cardinality of the observed variables? Pursuant to this objective, we were able to prove the nonclassicality of a variant of the Fritz distribution where $C$ has only two outcomes: the first corresponding to $A, B$ correlations and the second corresponding to $A, B$ anticorrelations\footnote{An inequality-free proof of the incompatibility of this distribution, using arguments analogous those presented in~\cite{Fritz_2012}, can be found in~\cite{Weilenmann_2016}. Assessing the compatibility of this variant using the methods in \cref{sec:marginal_satisfiability,sec:inflation_technique_summary,sec:deriving_inequalities} was first suggested to the authors by Denis Rosset.}. Nonetheless, such a distribution fails to deviate significantly from Bell nonclassicality.

    It is worth noting that there are quantum-accessible distributions which are conjectured, but not proven, to be incompatible with the Triangle structure. For instance, the distribution proposed in~\cite{Gisin_2017} required the use of entangled measurements and thus if proven incompatible might constitute a different form of quantum nonclassicality. Unfortunately, as was previously noted in~\cref{sec:found_inequalities}, the proposed distribution in~\cite{Gisin_2017} satisfies all inequalities generated by the inflations considered in~\cref{fig:inflations}. Presently, computational limitations prevent us from considering larger inflations than those in~\cref{fig:inflations}, although we remain optimistic.

    In truth, properly defining the novelty of quantum correlations and also recognizing the resourcefulness of those correlations in the Triangle structure, or any other causal structure, is a deep and meaningful, albeit unsettled, problem. Ultimately, if a satisfactory classifier of novelty is constructed, it remains unclear whether or not novel distributions even exist.

    \section{Conclusions}
    \label{sec:conclusions}
    In \cref{sec:triangle_structure}, we elucidated that establishing or rejecting compatibility with the Triangle structure has been a challenging problem for nearly a decade \cite{Steudel_2010,Branciard_2012,Henson_2014,Fritz_2012,Gisin_2017}. Though some causal compatibility inequalities were known, those inequalities did not appear useful for the purpose of witnessing quantum nonclassicality. Recently,~\citet{Fritz_2012} gave the first example of quantum nonclassicality in the Triangle structure, relying on an inequality-free proof that is not robust to any amount of noise.

    In \cref{sec:found_inequalities} we presented the first examples of causal compatibility inequalities capable of having quantum violations in the sense that they are violated by quantum-accessible distributions. This result was made possible through the Inflation Technique~\cite{Inflation} (\cref{sec:inflation_technique_summary}) applied to the Fritz distribution. Moreover, the inequalities in \cref{sec:found_inequalities} were derived using the inflations in \cref{fig:inflations}, each of which is low in the hierarchy proposed by~\cite{Navascues_2017}, thus revealing the relative efficiency of the Inflation Technique.

    In \cref{sec:noise} it was demonstrated that these causal compatibility inequalities are robust to noise, directly revealing a critical departure from Fritz's original proof of the incompatibility of the Fritz's perfect-correlation example. In \cref{sec:optimizations} we found quantum distributions \textit{quantitatively} distinct from the Fritz's recycled Bell theorem example, such that these optimized distributions more strongly violate certain causal compatibility inequalities.

    Despite these advancements, the distributions we discovered hew closely to the Fritz distribution, indicating that their nonclassical nature remains some recycled version of the nonclassicality found in the Bell structure. \Cref{sec:fritz_problem_revisit} discusses potential proposals certifying the genuineness of nonclassicality in the Triangle structure. Presently, the existence, and subsequent suitable classification of fundamentally novel nonclassicality remains speculative, and certainly warrants future research.

    \begin{acknowledgments}
    This research was supported in part by Perimeter Institute for Theoretical Physics. Research at Perimeter Institute is supported by the Government of Canada through the Department of Innovation, Science and Economic Development Canada and by the Province of Ontario through the Ministry of Research, Innovation and Science. The authors are grateful to Miguel Navascués and Denis Rosset for their insightful discussions. The authors would like to thank Rafael Chaves for identifying an error with the Wagon-Wheel inequality found in earlier versions of this paper.
    \end{acknowledgments}

    \appendix

    \section{Marginal Satisfiability and Inequalities}
    \label{sec:marginal_satisfiability}

    This section aims to explain how to solve the following decision problem: given a collection of probability distributions $\bc{\prob[V_1], \ldots, \prob[V_m]}$ where each set of variables $V_{i} \subseteq \jointvar$ is a subset of some complete set of variables $\jointvar$, does there exist a joint distribution $\prob[\jointvar]$ such that each $\prob[V_i]$ can be obtained by marginalizing $\prob[\jointvar]$ over the variables not in $V_i$, i.e. $\prob[V_i] = \sum_{\jointvar \setminus V_i} \prob[\jointvar]$? Colloquially, this problem is referred to as \term{the marginal problem}~\cite{Fritz_2011}. Additionally, this section aims to accomplish something further: If such as joint distribution $\prob[\jointvar]$ exists, how does one find it? If not, how does one find an inequality whose violation by $\bc{\prob[V_1], \ldots, \prob[V_m]}$ proves the nonexistence of a joint distribution. This is accomplished by illustrating how the marginal problem can be expressed as a linear program in which the solution to the marginal problem is encoded in the feasibility or infeasibility of said linear program. This section is presented prior to \cref{sec:inflation_technique_summary} as the marginal problem becomes an integral component of the Inflation Technique in subsequently deriving the inequalities presented in \cref{sec:found_inequalities}. Moreover, the Marginal problem is presented here logically independent from the remainder of the manuscript both for procedural clarity and because the marginal problem has applications to numerous areas of mathematics including game theory~\cite{Vorobev_1962}, database theory, knowledge integration of expert system, and of course, quantum information theory~\cite{Fritz_2011}.

    To begin, several pieces of nomenclature will be introduced to facilitate discussions. First, the set $\mscenario = \bc{V \mid V \subseteq \jointvar}$ of subsets of $\jointvar$ is referred to as the \term{marginal scenario} and each element $V \in \mscenario$ is termed a \term{(marginal) context} of $\mscenario$. The complete set of marginal distributions is referred to as the \term{marginal model} and is denoted with an superscript $\prob^{\mscenario} \defined \bc{\prob[V] \mid V \in \mscenario}$. A marginal model acts as the most general description of a family of observations that can be made over $\jointvar$. Strictly speaking, as defined by~\cite{Fritz_2011}, a marginal scenario forms an \textit{abstract simplicial complex} where it is required that all subsets of contexts are also contexts: $\forall V \in \mscenario, V' \subset V : V' \in \mscenario$. Throughout this section, we exclusively consider (without loss of generality) the \textit{maximal} marginal scenario; restricting our focus to the largest marginal contexts. Additionally, all marginal scenarios are taken to be \text{complete} in the sense that the marginal scenario covers the complete set of observable variables, i.e $\jointvar = \bigcup_{V \in \mscenario} V$. Finally, we henceforth assume that each variable $v \in \s J$ has a finite cardinality.

    The marginal problem asks: given a marginal model $\prob^{\mscenario} = \bc{\prob[V] \mid V \in \mscenario}$ marginal to the joint variables $\jointvar$, does there exist a joint distribution $\prob[\jointvar]$ such that each context $\prob[V]$ can be obtained by marginalizing $\prob[\jointvar]$?
    \[ \forall V \in \mscenario : \prob[V] = \sum_{\jointvar \setminus V} \prob[\jointvar] \eq \label{eq:def:marginal_problem}\]
    A marginal model $\prob^{\mscenario}$ is said to be \textit{contextual} if it \textit{does not} admit a joint distribution and noncontextual otherwise. Notice that \cref{eq:def:marginal_problem} is inherently a linear system of constraints which can be solved efficiently using linear programs. In consideration of this, we will now endeavor to discuss how to cast the \cref{eq:def:marginal_problem} as a matrix multiplication equation so that it becomes possible to discuss existing methods for deriving constraints on the set of contextual marginal models.

    To every discrete random variable $v$ there corresponds a prescribed set of \term{outcomes} $O_v$. We also define the set of all \term{events over $v$}, denoted $\Events{v}$,\footnote{In the language of sheaf theory, $\Events{v}$ is the \textit{sheaf of events}~\cite{Abramsky_2011}.} to be the set of all functions $s : \bc{v} \to O_v$ each representing the event that a measurement on $v$ was made where $s\br{v} \in O_v$ was observed. Evidently, $\Events{v}$ and $O_v$ have a one-to-one correspondence and this distinction can be confounding. There is rarely any harm in referring synonymously to either as outcomes. Nonetheless, a sheaf-theoretic treatment of contextuality~\cite{Abramsky_2011} demands the distinction.
    Specifically for this work, the distinction becomes essential for our discussion and exploit of marginal symmetries in \cref{sec:symmetric_inequalities}. As a natural generalization we define the events over a set of random variables $V = \bc{v_1, \ldots, v_n}$ in a parallel manner,
    \[ \Events{V} \defined \bc{s: V \to O_{V} \mid \forall i : s\br{v_i} \in O_{v_i} } \eq \label{eq:outcome_space}\]
    Each event $s$ can be compactly represented as a set of mappings over each element of $V$, i.e. $s = \bc{v_i\mapsto s\br{v_i}}_{i = 1}^{k}$. The domain $\Dom{s}$ of an event $s$ is the set of random variables it valuates, i.e. if $s \in \Events{V}$, then $\Dom{s} = V$. Under this framework, a probability distribution $\prob[V]$ can be considered as a map from $\Events{V}$ to the unit interval $\bs{0,1}$.
    The marginal problem inherently depends on the concept of probabilistic marginalization. This concept can be understood at the level of events; one event $s \in \Events{V}$ can be ``marginalized'' or restricted to a smaller event $s' \in \Events{W}$ whenever $W \subseteq V$. For every $W \subseteq V$ and $s \in \Events{V}$, the \term{restriction of $s$ onto $W$} (denoted $s|_{W} \in \Events{W}$) is the event in $\Events{W}$ that \textit{agrees} with each of $s$'s assignments for variables in $W$: $\forall v \in W : s|_{W}\br{v} = s\br{v}$.

    For every marginal scenario $\mscenario = \bc{V_1, \ldots, V_k}$, it is useful to put special emphasis on the \term{joint events} $\Events{\jointvar}$ which represent all possible global events over the entire set of joint variables. Similarly, we define the \term{context events} for a particular context $V \in \mscenario$ as $\Events{V}$. Finally, we elect to define the \term{marginal events} as the disjoint union over all context events and by an abuse of notation we will denote this union as $\Events{\mscenario} = \coprod_{V \in \mscenario} \Events{V}$. Each marginal section $m \in \Events{\mscenario}$ has a domain $\Dom{m} = V$ for some $V \in \mscenario$. By construction each marginal event $m \in \Events{\mscenario}$ is a restriction of some joint event $j \in \Events{\jointvar}$.

    The marginalization operation of the marginal problem is a \textit{linear} operation, mapping a joint probability distribution $\prob[\jointvar] : \Events{\jointvar} \to \bs{0,1}$ into a marginal model $\prob^{\mscenario} = \bc{\prob[V] : \Events{V} \to \bs{0,1} \mid V \in \mscenario}$. Since $\Events{\mscenario}$ and $\Events{\jointvar}$ are finite, the marginal problem can be represented as a $\abs{\Events{\mscenario}} \times \abs{\Events{\jointvar}}$ matrix.

    \begin{definition}
        \label{def:incidence_matrix}
        The \term{incidence matrix} $M$ for a marginal scenario $\mscenario = \bc{V_1, \dots, V_k}$ is a bitwise matrix where the columns are indexed by \textit{joint} events $j \in \Events{\jointvar}$ and the rows are events by \textit{marginal} events $m \in \Events{\mscenario}$. The entries of $M$ are populated whenever a marginal event $m$ is a restriction of the joint event $j$.
        \[ M_{m,j} \defined \begin{cases}
            1 & m = j|_{\s{D}\br{m}} \\
            0 & \text{otherwise}
        \end{cases} \eq \label{eq:incidence_definition}\]
        The incidence matrix has $\abs{\Events{\jointvar}}$ columns, $\abs{\Events{\mscenario}} = \sum_{i} \abs{\Events{V_i}}$ rows and $k\abs{\Events{\jointvar}}$ nonzero entries.
    \end{definition}
    To illustrate this concretely, consider the following example. Let $\jointvar$ be $3$ binary variables $\jointvar = \bc{A,B,C}$ and $\mscenario$ be the marginal scenario $\mscenario = \bc{\bc{A,B}, \bc{B,C}, \bc{A,C}}$. The incidence matrix for $\mscenario$ is:
    \[ M = \kbordermatrix{
        (A,B,C) \:\: \mapsto & (0,0,0) & (0,0,1) & (0,1,0) & (0,1,1) & (1,0,0) & (1,0,1) & (1,1,0) & (1,1,1) \\
        (A\mapsto0, B\mapsto0) & \kone & \kone & \kzer & \kzer & \kzer & \kzer & \kzer & \kzer \\
        (A\mapsto0, B\mapsto1) & \kzer & \kzer & \kone & \kone & \kzer & \kzer & \kzer & \kzer \\
        (A\mapsto1, B\mapsto0) & \kzer & \kzer & \kzer & \kzer & \kone & \kone & \kzer & \kzer \\
        (A\mapsto1, B\mapsto1) & \kzer & \kzer & \kzer & \kzer & \kzer & \kzer & \kone & \kone \\
        (B\mapsto0, C\mapsto0) & \kone & \kzer & \kzer & \kzer & \kone & \kzer & \kzer & \kzer \\
        (B\mapsto0, C\mapsto1) & \kzer & \kone & \kzer & \kzer & \kzer & \kone & \kzer & \kzer \\
        (B\mapsto1, C\mapsto0) & \kzer & \kzer & \kone & \kzer & \kzer & \kzer & \kone & \kzer \\
        (B\mapsto1, C\mapsto1) & \kzer & \kzer & \kzer & \kone & \kzer & \kzer & \kzer & \kone \\
        (A\mapsto0, C\mapsto0) & \kone & \kzer & \kone & \kzer & \kzer & \kzer & \kzer & \kzer \\
        (A\mapsto0, C\mapsto1) & \kzer & \kone & \kzer & \kone & \kzer & \kzer & \kzer & \kzer \\
        (A\mapsto1, C\mapsto0) & \kzer & \kzer & \kzer & \kzer & \kone & \kzer & \kone & \kzer \\
        (A\mapsto1, C\mapsto1) & \kzer & \kzer & \kzer & \kzer & \kzer & \kone & \kzer & \kone
    } \eq \label{eq:example_incidence}\]
    The incidence matrix acts (to the right) on a vector representing the joint probability $\prob[\jointvar]$ distribution and outputs a vector representing the marginal model $\prob^{\mscenario}$. The \term{joint distribution vector} $\probvec^{\jointvar}$ for a probability distribution $\prob[\jointvar]$ is the vector indexed by joint events $j \in \Events{\jointvar}$ whose entries are populated by the probabilities that $\prob[\jointvar]$ assigns to each joint event: $\probvec^{\jointvar}_j = \prob[\jointvar][j]$. Analogously, \term{marginal distribution vector} $\probvec^{\mscenario}$ for a marginal model $\prob^{\mscenario}$ is the vector whose entries are probabilities over the set of marginal outcomes $\Events{\mscenario}$: $\probvec^{\mscenario}_m = \prob[\Dom{m}][m]$.

    By design, the marginal and joint distribution vectors are related via the incidence matrix $M$. Given a joint distribution vector $\probvec^{\jointvar}$ one can obtain the marginal distribution vector $\probvec^{\mscenario}$ by multiplying $M$ by $\probvec^{\jointvar}$.
    \[ \probvec^{\mscenario} = M \cdot \probvec^{\jointvar} \eq \label{eq:incidence_matrix_use} \]
    As a quick remark, the particular ordering of the rows and columns of $M$ carries \textit{no importance}, but it \textit{must} be consistent between $M$, $\probvec^{\jointvar}$ and $\probvec^{\mscenario}$. The marginal problem can now be rephrased in the language of the incidence matrix. Suppose one obtains a marginal distribution vector $\probvec^{\mscenario}$; the marginal problem becomes equivalent to the question: does there exist a joint distribution vector $\probvec^{\jointvar}$ such that \cref{eq:incidence_matrix_use} holds? This question is naturally framed as the \term{marginal linear program}:
    \begin{equation}
    \begin{aligned}
        & \text{minimize:} \quad&& \emptyset \cdot x\\
        & \text{subject to:} && x \succeq 0 \\
        & && M \cdot x = \probvec^{\mscenario}
    \end{aligned}
    \end{equation}
    If this ``optimization''\footnote{``Optimization'' is presented in quotes here because the minimization objective is trivially always zero ($\emptyset$ denotes the null vector of all zero entries). The primal value of the linear program is of no interest, all that matters is its \textit{feasibility}.} is \textit{feasible}, then there exists a vector $x$ than can satisfy \cref{eq:incidence_matrix_use} and is a valid joint distribution vector. Therefore, feasibility of the marginal linear program not only implies that $\probvec^{\jointvar}$ exists but returns $\probvec^{\jointvar}$. Moreover if the marginal linear program is \textit{infeasible}, then there \textit{does not} exist a joint distribution $\probvec^{\jointvar}$. To every linear program, there exists a dual linear program that characterizes the feasibility of the original~\cite{Schrijver_1998}. Constructing the dual linear program is straightforward~\cite{Lahaie_2008}.
    \begin{equation}
    \begin{aligned}
        & \text{minimize:} \quad&& y \cdot \probvec^{\mscenario}\\
        & \text{subject to:} && y \cdot M \succeq 0
    \end{aligned}
    \end{equation}

    The dual marginal linear program not only answers the marginal problem for a specific marginal model $\probvec^{\mscenario}$ but as a by-product provides an inequality that witnesses its contextuality. If this is not obvious, first notice that the dual problem is \textit{never infeasible}; by choosing $y$ to be trivially the null vector $\emptyset$ of appropriate size, all constraints become satisfied. Secondly if the dual constraint $y \cdot M \succeq 0$ holds \textit{and} the primal is feasible, then $y \cdot \probvec^{\mscenario} =  y \cdot M \cdot x \geq 0$. Therefore the \textit{sign} of the dual objective $d \defined \min \br{y \cdot \probvec^{\mscenario}}$ classifies a marginal model's contextuality; if $d < 0$ then $y \cdot \probvec^{\mscenario} \geq 0$ is violated and therefore $\probvec^{\mscenario}$ is contextual. Likewise if $d \geq 0$ (satisfying $y \cdot \probvec^{\mscenario}$), then $\probvec^{\mscenario}$ is noncontextual.\footnote{Actually, if $d \geq 0$ then it is exactly $d = 0$ due to the existence of the trivial $y = \emptyset$. This observation is an instance of the \textit{Complementary Slackness Property} of~\cite{Bradley_1977}. Moreover, if $d < 0$, then it is unbounded $d = -\inf$. This latter point becomes clear upon recognizing that for any $y$ with $d < 0$, another $y' = \al y$ can be constructed (with $\al > 1$) such that $d' = \al d < d$. Since a more negative $d'$ can always be found, it must be that $d$ is unbounded. This is a demonstration of the fundamental \textit{Unboundedness Property} of~\cite{Bradley_1977}; if the dual is unbounded, then the primal is infeasible.} This is manifestation of Farkas's lemma~\cite{Schrijver_1998}. An \term{infeasibility certificate}~\cite{Andersen_2001} is any vector $y$ that satisfies the $y \cdot M \succeq 0$. Most linear program software packages such as Mosek~\cite{Mosek_2016}, Gurobi~\cite{Gurobi_2016}, CPLEX~\cite{Cplex_2016} and CVX/CVX OPT~\cite{CVX_2016,CVX_Opt_2016} are capable of producing infeasibility certificates. Furthermore, for every $y$ satisfying $y \cdot M \succeq 0$ there corresponds a \term{certificate inequality} that constraints the set of noncontextual marginal models. If $y$ is an infeasibility certificate, then $y \cdot \probvec^{\mscenario} \geq 0$ is satisfied by all contextual marginal models.

    The marginal problem can sometimes take on a more general variant that does not begin with a specific marginal model~\cite{Abramsky_2012,Mansfield_2012,Fritz_2011}: \textit{Given a marginal scenario $\mscenario$, what is the set of all noncontextual marginal models?} \citet{Pitowsky_1991} demonstrates that the set of noncontextual marginal models forms a \textit{convex} polytope called the \term{marginal polytope}. The extremal rays of a marginal polytope directly correspond to the columns of $M$ which further correspond to \textit{deterministic} joint distributions $\probvec^{\jointvar}$. Since all joint distributions $\probvec^{\jointvar}$ are probability distributions, their entries must sum to unity $\sum_{j \in \Events{\jointvar}} \probvec^{\jointvar}_j = 1$. This normalization defines the convexity of the polytope; all noncontextual marginal models are convex mixtures of the deterministic marginal models pursuant to \cref{eq:incidence_matrix_use}. The marginal polytope is a beneficial tool for understanding contextuality. First, the \textit{facets} of a marginal polytope correspond to a finite set of linear inequalities that are complete in the sense that all contextual distributions violate at least one facet inequality~\cite{Brunner_2013}. From the perspective of a marginal polytope, convex hull algorithms or linear quantifier elimination can be used to compute a representation of the complete set of linear inequalities and completely solve the marginal problem. A popular tool for linear quantifier elimination is \term{Fourier-Motzkin elimination} \cite{Dantzig_1973,Inflation,Abramsky_2012}. Applying the Fourier-Motzkin procedure to completely solve the marginal problem is discussed in more detail in \citet{Fritz_2011}. An excellent survey of existing techniques for solving the marginal problem including Equality Set Projection~\cite{jones_2004} and Hardy-type hypergraph transversals can be found in \citet{Inflation}. In conclusion, there are a number of computational tools available to solve the marginal problem completely whenever no marginal model is provided.

    Each of the above mentioned techniques suffers from computational complexity limitations. For example, the Fourier-Motzkin procedure is in the worst case doubly exponential in the number of initial inequalities~\cite{Dantzig_1973}. For the purposes of this research, solving the marginal problem without reference to a marginal model was intractable. This will become apparent in \cref{sec:deriving_inequalities} when the Inflation Technique is applied to the Triangle structure, producing considerably large marginal models. Luckily, the Fritz distribution allows one to avoid the complexity issues of the complete marginal problem and instead focus on the original problem of determining whether or not a particular marginal model admits a joint distribution or not.

    \section{Inflation Technique}
    \label{sec:inflation_technique_summary}
    The \term{Inflation Technique}, invented by \citet{Inflation} and inspired by the \textit{do calculus} and \textit{twin networks} of \citet{Pearl_2009}, is a family of causal inference techniques that can be used to determine if an observable probability distribution $\prob[\nodes_{O}]$ is compatible or incompatible with a given causal structure $\graph$. As a precursor, the Inflation Technique begins by \textit{augmenting} a causal structure $\graph$ with additional copies of its nodes, producing an \textit{inflated} causal structure $\graph'$ called an \term{inflation}, and then exposes how causal inference tasks on the inflation can be used to make inferences on the original causal structure. For reference, a few inflations of the Triangle structure are depicted in \cref{fig:inflations}. Copies of nodes in the inflated causal structure are distinguished by an additional subscript called the \term{copy-index}. For example, node $A$ of \cref{fig:triangle_structure} has copies $A_1, A_2, A_3, A_4$ in the inflated causal structure in \cref{fig:the_web_inflation}. All such copies are deemed equivalent via a \term{copy-index equivalence} relation denoted `$\sim$'. A copy-index is effectively arbitrary, so we will refer to an arbitrary inflated copy of $A$ as $A'$, i.e. $A \sim A_1 \sim A' \not\sim B \sim B_1 \sim B'$.\footnote{Note that we preemptively generalize the notion of copy-index equivalence to other mathematical objects like sets, graphs, and groups by saying that $X \sim Y$ if and only if $X$ is equivalent to $Y$ upon removal of the copy-index.}

    \begin{nscenter}
    \begin{figure}
    \begin{subfigure}[b]{.30\linewidth}
    \scalebox{0.65}{\includegraphics{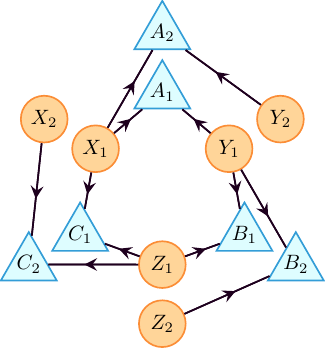}}
    \caption{}
    \label{fig:spiral_inflation}
    \end{subfigure}
    \begin{subfigure}[b]{.30\linewidth}
    \scalebox{0.65}{\includegraphics{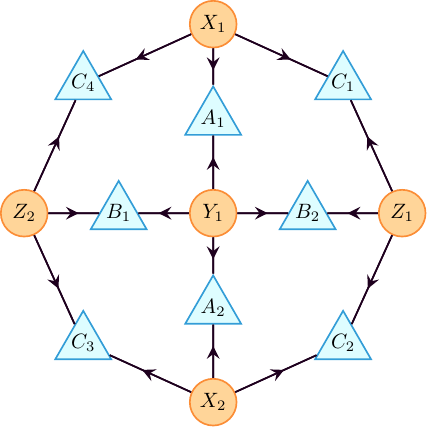}}
    \caption{}
    \label{fig:wagon_wheel_inflation}
    \end{subfigure}
    \begin{subfigure}[b]{.30\linewidth}
    \scalebox{0.65}{\includegraphics{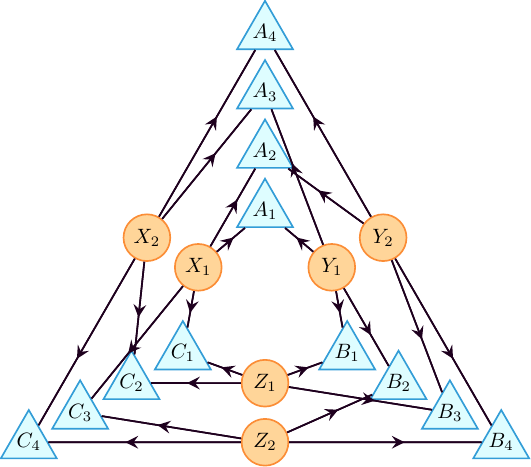}}
    \caption{}
    \label{fig:the_web_inflation}
    \end{subfigure}
    \caption{Some inflations of the Triangle structure: (a) the Spiral inflation, (b) the Wagon-Wheel inflation, and (c) the Web Inflation.}
    \label{fig:inflations}
    \end{figure}
    \end{nscenter}

    In addition to the common graph-theoretic terminology and notation presented in \cref{sec:causal_compatibility}, two related concepts need introductions. First, an \term{induced subgraph} of $\graph$ for a subset of nodes $N \subseteq \nodes$ is the graph composed of nodes $N$ and all edges $e$ of the original graph that are contained in $N$: $\Sub[\graph]{N} \defined \br{N, \bc{e = \bc{n \to m} \mid n, m \in N}}$. An \term{ancestral subgraph} of $\graph$ for a subset of nodes $N \subseteq \nodes$ is the induced subgraph due to the ancestry of $N$: $\AnSub[\graph]{N} \defined \Sub[\graph]{\An[\graph]{N}}$.

    The Inflation Technique begins with distribution $\prob[N]$ defined over some observable nodes $N$ of $\graph$ and the a priori assumption that it is compatible with $\graph$ pursuant to the definitions provided in \cref{sec:causal_compatibility}. Regarding $\graph$ as a causal hypothesis, the observable correlation $\prob[N]$ can only be influenced by the ancestry of $N$ in $\graph$. Consequently, for any set of nodes $N'$ of an inflation $\graph'$ where the ancestral subgraph $\AnSub[\graph']{N'}$ happens to be homomorphic to the ancestral subgraph $\AnSub[\graph]{N}$, one can conclude that the distribution $\prob[N']$ (induced by $\prob[N]$\footnote{The inflated distribution $\prob[N']$ assigns the same probability to all events of $\prob[N]$ whenever the events are equivalent under the removal of copy-indices.}) is compatible with $\graph'$ using the same latent explanations for $\prob[N]$ in $\graph$. This observation is known as the \term{Inflation Lemma}~\cite[Lemma 3]{Inflation}.

    To formalize the Inflation Lemma, we define the \term{injectable sets of $\graph'$}, denoted $\Inj[\graph]{\graph'}$, as all sets of nodes in $\graph'$ whose ancestral subgraphs are homomorphic (via copy-index removal) to an ancestral subgraph in $\graph$: $\Inj[\graph]{\graph'} \defined \bc{N' \subseteq \nodes' \mid \exists N \subseteq \nodes : \AnSub[\graph']{N'} \sim \AnSub[\graph]{N}}$. Analogously defined are the \term{images of the injectable sets in $\graph$}: $\ImInj[\graph]{\graph'} \defined \bc{N \subseteq \nodes \mid \exists N' \subseteq \nodes' : \AnSub[\graph']{N'} \sim \AnSub[\graph]{N}}$.

    To illustrate these concepts, consider the Spiral inflation $\graph'$ depicted in \cref{fig:spiral_inflation}. The ancestral subgraph of $\bc{A_1}$ in the Spiral inflation (denoted $\AnSub[\graph']{\bc{A_1}}$) is highlighted in \cref{fig:spiral_inflation_inj1} and is clearly homomorphic to the ancestral subgraph of its image $\bc{A}$ in the Triangle structure ($\graph$) (\cref{fig:spiral_inflation_iminj1}). Additionally, the set $\bc{A_2, C_1}$ is an injectable set of $\graph'$ because $\AnSub[\graph']{\bc{A_2, C_1}}$ (highlighted in \cref{fig:spiral_inflation_inj2}) is homomorphic (via copy-index removal) to the $\AnSub[\graph]{\bc{A, C}}$ in the Triangle structure $(\graph)$ (\cref{fig:spiral_inflation_iminj2}).

    \begin{nscenter}
    \begin{figure}
    \begin{subfigure}[b]{.23\linewidth}
    \scalebox{0.8}{\includegraphics{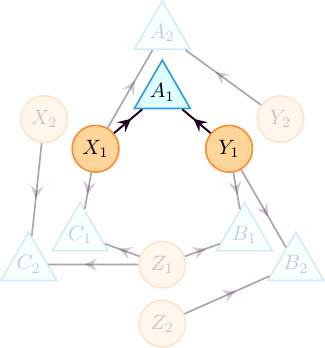}}
    \caption{}
    \label{fig:spiral_inflation_inj1}
    \end{subfigure}
    \begin{subfigure}[b]{.23\linewidth}
    \scalebox{0.8}{\includegraphics{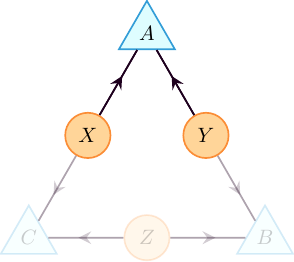}}
    \caption{}
    \label{fig:spiral_inflation_iminj1}
    \end{subfigure}
    \begin{subfigure}[b]{.23\linewidth}
    \scalebox{0.8}{\includegraphics{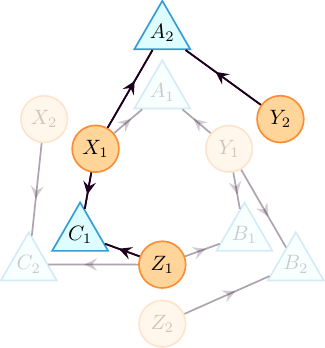}}
    \caption{}
    \label{fig:spiral_inflation_inj2}
    \end{subfigure}
    \begin{subfigure}[b]{.23\linewidth}
    \scalebox{0.8}{\includegraphics{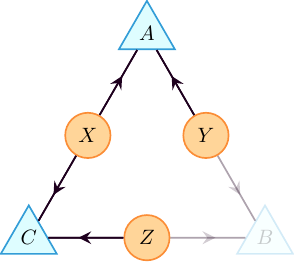}}
    \caption{}
    \label{fig:spiral_inflation_iminj2}
    \end{subfigure}
    \caption{Some injectable sets $\Inj[\graph]{\graph'}$ of the Spiral inflation ($\graph'$) and their corresponding images $\ImInj[\graph]{\graph'}$ in the Triangle structure ($\graph$): (a) $\AnSub[\graph']{\bc{A_1}}$, (b) $\AnSub[\graph]{\bc{A}}$, (c) $\AnSub[\graph']{\bc{A_2, C_1}}$, and (d) $\AnSub[\graph]{\bc{A,C}}$.}
    \label{fig:injectablesets}
    \end{figure}
    \end{nscenter}

    The injectable sets of an inflation are of \textit{principle importance} to the Inflation Technique. Given a distribution $\prob[\nodes_{O}]$ defined over the observable nodes of $\graph$, one can compute a marginal model defined over the injectable sets $\prob^{\Inj[\graph]{\graph'}} = \bc{\prob[N'] \mid N' \in \Inj[\graph]{\graph'}}$ and their images $\prob^{\ImInj[\graph]{\graph'}} = \bc{\prob[N] \mid N \in \ImInj[\graph]{\graph'}}$. The contrapositive of the Inflation Lemma casts the compatibility problem for $\prob^{\ImInj[\graph]{\graph'}}$ (which exhibits equivalent compatibility as $\prob[\nodes_O]$) into a compatibility problem for $\prob^{\Inj[\graph]{\graph'}}$ of any nontrivial inequalities. If $\prob^{\Inj[\graph]{\graph'}}$ is found to be incompatible with $\graph'$, then $\prob^{\ImInj[\graph]{\graph}}$ (and also $\prob[\nodes_O]$) must be incompatible with $\graph$. Fortunately, the inflated causal structure $\graph'$ possesses its own \textit{d-separation relations} which enforces conditional independence equality constraints on $\prob^{\Inj[\graph]{\graph'}}$. A useful subset of d-separation relations are those corresponding to unconditional d-separations, known as \term{ancestrally independent} sets~\cite{Pearl_2009}. Two sets $N_1', N_2'$ are ancestrally independent ($N_1' \ancestralindep N_2'$) if they have distinct ancestry in $\graph'$.
    \[ N_1' \ancestralindep N_2' \iff \An[\graph']{N_1'} \cap \An[\graph']{N_2'}= \emptyset \eq \]
    Ancestral independence implies an unconditional probabilistic independence: if $N_1' \ancestralindep N_2'$ then $\prob[N_1' \cup N_2'] = \prob[N_1']\prob[N_2']$. If $N_1', N_2' \in \Inj[\graph]{\graph'}$, then the associated probabilistic constraint is applicable to $\prob^{\Inj[\graph]{\graph'}}$. This notion generalizes to more than two ancestrally independent sets. A set $N' \subseteq \nodes'$ is an \term{ai-expressible set} if it can be decomposed into the disjoint union of injectable sets $N' = \coprod_{i} N'_i \mid N'_i \in \Inj[\graph]{\graph'}$ and all pairs $N_i', N_j'$ are ancestrally independent: $\forall i, j : N_i' \ancestralindep N_j'$. The analogous probabilistic constraint is: $\prob[N'] = \prob[\coprod_i N'_i] = \prod_{i}\prob[N'_i]$~\cite{Pearl_2009}.\footnote{In~\cite{Inflation}, this concept is generalized into terms which can be factorized via d-separation conditions and the corresponding inflated sets are termed \textit{expressible sets}. This generalization was not required for this work in particular because the ancestral independence relations formed a generating set of d-separation conditions for all of the inflations considered in \cref{fig:inflations}.} Throughout this work, we let $\AIExpr[\graph]{\graph'}$ denote the set of all ai-expressible sets.\footnote{Analogously to a marginal scenario $\mscenario$, the ai-expressible sets $\AIExpr[\graph]{\graph'}$ form an \textit{abstract simplicial complex}. Therefore, in practice, it is completely sufficient to focus on the \textit{maximal} ai-expressible sets.} Efficient algorithms for computing the injectable and ai-expressible sets of an inflation can be found in~\cite{Inflation}. \Cref{tab:asexpr_triangle_structure_wagon_wheel,tab:asexpr_triangle_structure_web} tabulate the ai-expressible sets and the associated ancestral separations for the Wagon-Wheel and Web inflations (respectively \cref{fig:wagon_wheel_inflation,fig:the_web_inflation}).

    \begin{nscenter}
        \begin{table}
        \parbox{.45\linewidth}{
            \centering
            \begin{tabular}{|c|c|}
                \hline
                \multicolumn{2}{|c|}{Maximally AI-Expressible Sets} \\
                \multicolumn{2}{|c|}{for the Wagon-Wheel Inflation} \\
                \hline
                $\AIExpr[\graph]{\graph'}$ & Ancestral Separations \\
                \hline
                $\bc{A_2, B_1, C_3, C_1}$ & $\bc{A_2, B_1, C_3} \ancestralindep \bc{C_1}$ \\
                $\bc{A_1, B_1, C_4, C_2}$ & $\bc{A_1, B_1, C_4} \ancestralindep \bc{C_2}$ \\
                $\bc{A_1, B_2, C_1, C_3}$ & $\bc{A_1, B_2, C_1} \ancestralindep \bc{C_3}$ \\
                $\bc{A_2, B_2, C_2, C_4}$ & $\bc{A_2, B_2, C_2} \ancestralindep \bc{C_4}$ \\
                \hline
            \end{tabular}
            \caption{The maximally ai-expressible sets for the Wagon-Wheel Inflation.
            \label{tab:asexpr_triangle_structure_wagon_wheel}}
        }
        \hfill
        \parbox{.45\linewidth}{
            \centering
            \begin{tabular}{|c|c|}
                \hline
                \multicolumn{2}{|c|}{Maximally AI-Expressible Sets} \\
                \multicolumn{2}{|c|}{for the Web Inflation} \\
                \hline
                $\AIExpr[\graph]{\graph'}$ & Ancestral Separations \\
                \hline
                $\bc{A_1, B_1, C_1, A_4, B_4, C_4}$ & $\bc{A_1, B_1, C_1} \ancestralindep \bc{A_4, B_4, C_4}$ \\
                $\bc{A_1, B_2, C_3, A_4, B_3, C_2}$ & $\bc{A_1, B_2, C_3} \ancestralindep \bc{A_4, B_3, C_2}$ \\
                $\bc{A_2, B_3, C_1, A_3, B_2, C_4}$ & $\bc{A_2, B_3, C_1} \ancestralindep \bc{A_3, B_2, C_4}$ \\
                $\bc{A_2, B_4, C_3, A_3, B_1, C_2}$ & $\bc{A_2, B_4, C_3} \ancestralindep \bc{A_3, B_1, C_2}$ \\
                $\bc{A_1, B_3, C_4}$ & $\bc{A_1} \ancestralindep \bc{B_3} \ancestralindep \bc{C_4}$ \\
                $\bc{A_1, B_4, C_2}$ & $\bc{A_1} \ancestralindep \bc{B_4} \ancestralindep \bc{C_2}$ \\
                $\bc{A_2, B_1, C_4}$ & $\bc{A_2} \ancestralindep \bc{B_1} \ancestralindep \bc{C_4}$ \\
                $\bc{A_2, B_2, C_2}$ & $\bc{A_2} \ancestralindep \bc{B_2} \ancestralindep \bc{C_2}$ \\
                $\bc{A_3, B_3, C_3}$ & $\bc{A_3} \ancestralindep \bc{B_3} \ancestralindep \bc{C_3}$ \\
                $\bc{A_3, B_4, C_1}$ & $\bc{A_3} \ancestralindep \bc{B_4} \ancestralindep \bc{C_1}$ \\
                $\bc{A_4, B_1, C_3}$ & $\bc{A_4} \ancestralindep \bc{B_1} \ancestralindep \bc{C_3}$ \\
                $\bc{A_4, B_2, C_1}$ & $\bc{A_4} \ancestralindep \bc{B_2} \ancestralindep \bc{C_1}$ \\
                \hline
            \end{tabular}
            \caption{The maximally ai-expressible sets for the Web Inflation.
            \label{tab:asexpr_triangle_structure_web}}
        }
        \end{table}
    \end{nscenter}

    Unlike $\prob^{\ImInj[\graph]{\graph'}}$, which is noncontextual by construction, $\prob^{\Inj[\graph]{\graph'}}$ contains overlapping marginals meaning its contextuality remains unknown and must be determined using any of the techniques discussed in~\cref{sec:marginal_satisfiability}. More importantly, the Inflation Technique introduces constraints relating to the ai-expressible sets. In practice, the equality constraints implied by $\AIExpr[\graph]{\graph'}$ permits one to construct a marginal model defined over the ai-expressible sets $\AIExpr[\graph]{\graph'}$ resulting in greater resolution in classifying the compatibility of $\prob[\nodes_O]$. In summary, the Inflation Technique partially transforms the compatibility problem into a marginal problem, wherein one can solve the marginal problem (\cref{sec:marginal_satisfiability}) to either determine the compatibility of an observable distribution $\prob[\nodes_O]$ with $\graph$ or derive compatibility inequalities for $\graph$, the latter of which is discussed in \cref{sec:deriving_inequalities}.

    \section{Deriving Inequalities from the Inflation Technique}
    \label{sec:deriving_inequalities}

    \Cref{sec:inflation_technique_summary} summarizes how the Inflation Technique~\cite{Inflation} can cast the compatibility problem into the marginal problem by leveraging inflations of the Triangle structure. Explicitly, for a given inflation $\graph'$ of the Triangle structure, one constructs a marginal problem (\cref{sec:marginal_satisfiability}) for the marginal scenario $\mscenario$ composed of the maximal ai-expressible sets of $\graph'$, i.e. $\mscenario = \AIExpr[\graph]{\graph'}$.

    Recall that the first inequality (\cref{eq:ww_ineq}) presented in \cref{sec:found_inequalities} was derived using the Wagon-Wheel inflation (\cref{fig:wagon_wheel_inflation}). The Wagon-Wheel inflation possesses $4$ copies of $C$ ($C_1, C_2, C_3, C_4$) and $2$ copies of $A$ ($A_1, A_2$) and $B$ ($B_1, B_2$) arranged in the shape of a Wagon-Wheel. The maximal ai-expressible sets of the Wagon-Wheel inflation along with their ancestral dependences can be found in \cref{tab:asexpr_triangle_structure_wagon_wheel}. These maximal ai-expressible sets define a marginal scenario where the joint variables $\jointvar$ are the set of observable nodes in the Wagon-Wheel inflation $\jointvar = \nodes'_O$:
    \[ \mscenario = \AIExpr[\graph]{\graph'} = \bc{\bc{A_2, B_1, C_3, C_1}, \bc{A_1, B_1, C_4, C_2}, \bc{A_1, B_2, C_1, C_3}, \bc{A_2, B_2, C_2, C_4}} \eq \label{eq:wagon_wheel_marginal_model}\]
    Given that the Fritz distribution is defined over four-outcome variables, the variables in the marginal scenario are assigned four-outcomes as well. This marginal scenario $\mscenario$ then defines an incidence matrix $M$ capable of accommodating the Fritz distribution that has $\abs{\Events{\mscenario}} = 4 \cdot 4^{4} = 1024$ rows and $\abs{\Events{\jointvar}} = 4^8 = 65,536$ columns; this matrix is not reproduced here. The sheer size of the incidence matrix used here makes complete solutions of the marginal problem using tools such as linear quantifier elimination intractable. To construct the marginal distribution vector $\probvec^{\mscenario}$ for the Fritz distribution, one begins with $\probvec^{\mscenario}$ defined in a \textit{symbolic} form,
    \[ {\probvec^{\mscenario}}^{\intercal} = \underbrace{\br{\prob[A_2B_1C_3C_1]\br{0000}, \ldots, \prob[A_1B_1C_4C_2]\br{3232}, \ldots, \prob[A_2B_2C_2C_4]\br{3333}}}_{1024 \text{ entries}} \eq \label{eq:marginal_vector_wagon_wheel} \]
    Which can be factored into the ancestrally independent injectable sets using \cref{tab:asexpr_triangle_structure_wagon_wheel},
    \[ {\probvec^{\mscenario}}^{\intercal} = \br{\prob[A_2B_1C_3]\br{000}\prob[C_1]\br{0}, \ldots, \prob[A_1B_1C_4]\br{323}\prob[C_2]\br{2}, \ldots, \prob[A_2B_2C_2]\br{333}\prob[C_4]\br{3}} \eq \label{eq:marginal_vector_wagon_wheel_factors}\]
    And penultimately, each probability distribution in \cref{eq:marginal_vector_wagon_wheel_factors} is deflated by dropping copy-indices.
    \[ {\probvec^{\mscenario}}^{\intercal} = \br{\prob[ABC]\br{000}\prob[C]\br{0}, \ldots, \prob[ABC]\br{323}\prob[C]\br{2}, \ldots, \prob[ABC]\br{333}\prob[C]\br{3}} \eq \label{eq:marginal_vector_wagon_wheel_factors_dropped}\]
    This step is permitted because all of the remaining distributions in \cref{eq:marginal_vector_wagon_wheel_factors} are defined over the injectable sets of the Wagon-Wheel inflation. Finally, each of the elements of $\probvec^{\mscenario}$ are replaced with numerical values pursuant to the Fritz distribution. For example, $\prob[ABC]\br{323}\prob[C]\br{2}$ is assigned the following numerical value:
    \[ \prob[ABC]\br{323}\prob[C]\br{2} = \f{1}{32}\br{2 + \sqrt{2}} \cdot \f{1}{4} \simeq 0.0267 \eq \]
    The same is applied to all other entries of $\probvec^{\mscenario}$. Finally, this numerical version of $\probvec^{\mscenario}$ and $M$ are subjected to linear programing software and an infeasibility certificate $y$ was obtained, corresponding precisely to \cref{eq:ww_ineq} using \cref{eq:marginal_vector_wagon_wheel_factors_dropped} in bitwise notation.

    The remaining inequalities in \cref{sec:found_inequalities} were derived using the Web inflation of \cref{fig:the_web_inflation}. The maximal ai-expressible sets of the Wagon-Wheel inflation along with their ancestral dependences can be found in \cref{tab:asexpr_triangle_structure_web}. Analogously to the Wagon-Wheel inflation, these ai-expressible sets form a marginal scenario $\mscenario$ which defines an incidence matrix $M$ that has $\abs{\Events{\mscenario}} = 4 \cdot 4^{6} + 8 \cdot 4^3 = 16,896$ rows, $\abs{\Events{\jointvar}} = 4^{12} = 16,777,216$ columns and $201,326,592$ nonzero entires.

    \section{Deriving Symmetric Causal Compatibility Inequalities}
    \label{sec:symmetric_inequalities}

    \Cref{sec:inflation_technique_summary} detailed how to obtain causal compatibility inequalities for any causal structure by constructing a corresponding marginal problem (as defined in \cref{sec:marginal_satisfiability}) and supplying an incompatible distribution to generate an infeasibility certificate. \cref{eq:symmetric_web_inequality} presented the causal compatibility inequality $I\tsb{SymmetricWeb}$ which is symmetric under permutations of the variables $A,B,C$. This section aims to describe a general technique that can be used to derive  $I\tsb{SymmetricWeb}$ and other inequalities also exhibiting this symmetry. In brief, this is accomplished by grouping marginal events $m \in \Events{\mscenario}$ of a marginal scenario $\mscenario$ into orbits under the action of variable permutations.

    Exploiting symmetries of the marginal scenario is useful for a few distinct reasons. First, \citet{Bancal_2010} discuss computational advantages in considering symmetric versions of marginal polytopes mentioned in \cref{sec:marginal_satisfiability}; the number of extremal points typically grows exponentially in $\jointvar$, but only polynomial for the symmetric polytope. They also note a number of interesting inequalities (such as CHSH~\cite{CHSH_Original}) can be written in a way that is symmetric under the exchange of parties, demonstrating that nontrivial inequalities can be recovered from facets of a symmetric polytope. Secondly, numerical optimizations against symmetric inequalities lead to one of two interesting cases: either the extremal distribution is symmetric itself or it is not. The latter case generates a family of incompatible distributions obtained by applying symmetry operations on the extremal distribution.\footnote{If the extremal distribution happens to be asymmetric, then one can conclude the space of accessible distributions is nonconvex.}

    To clarify which symmetries we have in mind, first consider the marginal scenario $\mscenario = \bc{\bc{A, B, C}, \bc{C, D}, \bc{A, D}}$ where each variable in $\jointvar = \bc{A,B,C,D}$ has binary outcomes $\bc{0,1}$. ${I_{\mscenario} \defined \bc{\prob[ABC][010] \leq \prob[CD][00] + \prob[AD][01]}}$ is an inequality constraining the set of noncontextual marginal models $\prob^{\mscenario}$. Now, the contextuality of a distribution $\prob[ABCD]$ should be \textit{invariant} under those permutations of the variable labels in $\jointvar$  which map the marginal scenario to itself; but not all variable relabellings preserve $\mscenario$. An example of a permutation $\varphi \in \Perm{\jointvar}$\footnote{The permutation group $\Perm{S}$ over a set $S$ is the set of all bijective maps $\varphi : S \to S$.} which does \emph{not} preserve $\mscenario$ is $\varphi\bs{\bc{a,b,c,d}} = \bc{c,a,d,b}$. The action of $\varphi$ on $I_{\mscenario}$ is
    \begin{align*}
    \eq \label{eq:symmetry_example}
    \begin{split}
        \varphi\bs{I_{\mscenario}} &= \bc{\prob[\varphi\bs{abc}][010] \leq \prob[\varphi\bs{cd}][00] + \prob[\varphi\bs{ad}][01]} \\
        &= \bc{\prob[cad][010] \leq \prob[db][00] + \prob[cb][01]} \\
        &= \bc{\prob[acd][100] \leq \prob[bd][00] + \prob[bc][10]}
    \end{split}
    \end{align*}
    which yields a valid albeit irrelevant inequality, as the resulting inequality no longer pertains to  $\mscenario$. 

    Permutations $\varphi$ that modify the marginal scenario have no application within the framework of the Inflation Technique (\cref{sec:inflation_technique_summary}) because the Inflation Lemma only holds when the inflated inequality constrains injectable sets. Therefore, the desired set of symmetries for our purposes is a subgroup of $\Perm{\jointvar}$ that takes $\mscenario$ to $\mscenario$.

    The \term{variable permutation group} $\gp\br{\mscenario}$ for a marginal scenario $\mscenario$ is the joint permutation subgroup that stabilizes the marginal scenario, $\Phi\br{\mscenario} \defined \bc{\varphi \in \Perm{\jointvar} \mid \forall V \in \mscenario : \varphi\bs{V} \in \mscenario}$.
    In general, the variable permutation group $\gp\br{\mscenario}$ can be obtained using group stabilizer algorithms. After obtaining $\gp\br{\mscenario}$, one can take known compatibility inequalities $I_{\mscenario}$ can create a whole family of inequalities $\bc{\varphi\bs{I_{\mscenario}} \mid \varphi \in \gp\br{\mscenario}}$ that are valid for the same marginal scenario $\varphi\bs{I_{\mscenario}} = \varphi\bs{I}_{\varphi\bs{\mscenario}} = \varphi\bs{I}_{\mscenario}$.

    Although useful for reducing computational complexity~\cite{Bancal_2010}, we divert our attention to finding \textit{symmetric} inequalities, i.e. those where $\forall \varphi \in \gp\br{\mscenario} :  \varphi\bs{I_{\mscenario}} = I_{\mscenario}$. To generate inequalities that exhibit certain symmetries using the methods described in \cref{sec:marginal_satisfiability}, it is sufficient to perform a change of basis on the incidence matrix $M$ for a given marginal scenario $\mscenario$. Through repeated action of $\varphi \in \gp$ on marginal outcomes $m \in \Events{\mscenario}$ and joint outcomes $j \in \Events{\jointvar}$, one can define \term{group orbits} of $\gp$ in $\Events{\mscenario}$ and $\Events{\jointvar}$; respectively $\Orb[\gp]{m} \defined \bc{\varphi\bs{m} \mid \varphi \in \gp}$, $\Orb[\gp]{j} \defined \bc{\varphi\bs{j} \mid \varphi \in \gp}$. The action of $\varphi \in \gp$ on any outcome $f \in \Events{V}$ (denoted $\geo\bs{f}$) is defined as $\br{\varphi\bs{f}}\br{v} \defined f\br{\varphi^{-1}\bs{v}}$ pursuant to intuitive action used in \cref{eq:symmetry_example}. Using these group orbits, it is possible to contract the incidence matrix $M$ of a marginal scenario into a symmetrized version. The \term{symmetric incidence matrix} $M_{\gp}$ for a marginal scenario $\mscenario$ and the variable permutation group $\gp$ is a contracted version of the incidence matrix $M$ for $\mscenario$. Each row of $M_{\gp}$ corresponds to a marginal orbit $\Orb[\gp]{m}$. Analogously each column of $M_{\gp}$ corresponds to a joint orbit $\Orb[\gp]{j}$. The entries of $M_{\gp}$ are integers and correspond to summing over the rows and columns of $M$ that belong to each orbit.
    \[ \br{M_{\gp}}_{\Orb[\gp]{m}, \Orb[\gp]{j}} = \sum_{\substack{j' \in \Orb[\gp]{j} \\ m' \in \Orb[\gp]{m}}} M_{m',j'} \eq \]
    It is possible to analogously define a \term{symmetric joint distribution vector} $\probvec^{\jointvar}_{\Phi}$ indexed by $\Orb[\gp]{j}$, $\br{\probvec^{\jointvar}_{\Phi}}_{\Orb[\gp]{j}} = \sum_{j' \in \Orb[\gp]{j}} \probvec^{\jointvar}_{j'}$ and a \term{symmetric marginal distribution vector} $\probvec^{\mscenario}_{\Phi}$ indexed by $\Orb[\gp]{m}$, $\br{\probvec^{\mscenario}_{\Phi}}_{\Orb[\gp]{m}} = \sum_{m' \in \Orb[\gp]{m}} \probvec^{\mscenario}_{m'}$. Together, $M_{\gp}, \probvec^{\jointvar}_{\Phi}$ and $\probvec^{\mscenario}_{\Phi}$ define a new, \term{symmetric marginal problem}: $\probvec^{\mscenario}_{\Phi} = M_{\gp} \cdot \probvec^{\jointvar}_{\Phi}$. The symmetric marginal problem can be solved using the same computational methods discussed in \cref{sec:marginal_satisfiability} and will produce symmetric inequalities.

    In the context of the Inflation Technique, a variable symmetry $\Phi\br{\mscenario'}$ over an inflated marginal scenario $\mscenario'$ does not always correspond to a variable symmetry under deflation $\Phi\br{\mscenario}$. In order to derive deflated inequalities that are symmetric under an exchange of parties, it is required that $\Phi\br{\mscenario'} \sim \Phi\br{\mscenario}$ are equivalent up to copy-index.

    For the Triangle structure in particular, the variable permutation group is the set of permutations on $A, B, C$: $\Phi\br{\mscenario} = \Perm{A, B, C}$. For the Web inflation (\cref{fig:the_web_inflation}), we have obtained $\gp\br{\AIExpr[\graph]{\graph'}}$, an order $48$ group with the following $4$ generators:
    \begin{equation}
    \begin{aligned}[c]
    &\hspace{1mm}\varphi_1 \\
    A_1 &\to A_4 \\A_2 &\to A_3 \\A_3 &\to A_2 \\A_4 &\to A_1 \\B_1 &\to B_4 \\B_2 &\to B_3 \\B_3 &\to B_2 \\B_4 &\to B_1 \\C_1 &\to C_4 \\C_2 &\to C_3 \\C_3 &\to C_2 \\C_4 &\to C_1
    \end{aligned}
    \quad
    \begin{aligned}[c]
    &\hspace{1mm}\varphi_2 \\
    A_1 &\to A_1 \\A_2 &\to A_3 \\A_3 &\to A_2 \\A_4 &\to A_4 \\B_1 &\to C_1 \\B_2 &\to C_3 \\B_3 &\to C_2 \\B_4 &\to C_4 \\C_1 &\to B_1 \\C_2 &\to B_3 \\C_3 &\to B_2 \\C_4 &\to B_4
    \end{aligned}
    \quad
    \begin{aligned}[c]
    &\hspace{1mm}\varphi_3 \\
    A_1 &\to C_1 \\A_2 &\to C_2 \\A_3 &\to C_3 \\A_4 &\to C_4 \\B_1 &\to A_1 \\B_2 &\to A_2 \\B_3 &\to A_3 \\B_4 &\to A_4 \\C_1 &\to B_1 \\C_2 &\to B_2 \\C_3 &\to B_3 \\C_4 &\to B_4
    \end{aligned}
    \quad
    \begin{aligned}[c]
    &\hspace{1mm}\varphi_4 \\
    A_1 &\to A_1 \\A_2 &\to A_2 \\A_3 &\to A_3 \\A_4 &\to A_4 \\B_1 &\to B_2 \\B_2 &\to B_1 \\B_3 &\to B_4 \\B_4 &\to B_3 \\C_1 &\to C_3 \\C_2 &\to C_4 \\C_3 &\to C_1 \\C_4 &\to C_2
    \end{aligned}
    \end{equation}
    Notice that $\varphi_1, \varphi_2, \varphi_3, \varphi_4$ are all automorphisms of the web inflation. Moreover they stabilize the ai-expressible sets of \cref{tab:asexpr_triangle_structure_web}. Importantly,
    \[ \gp\br{\AIExpr[\graph]{\graph'}} \sim \Perm{A,B,C} \eq \]
    To see this, $\varphi_1$ and $\varphi_4$ become the identity element in $\Perm{A,B,C}$ upon removal of the copy-index, leaving $\varphi_2$ to generate reflections and $\varphi_3$ to generate rotations.

    The symmetric incidence matrix $M_{\Phi}$ for the Web inflation is considerably smaller than $M$. The number of rows of $M_{\Phi}$ is a number of distinct orbits $\Orb[\Phi]{m}$ in $\Events{\mscenario}$. Likewise the number of columns is the number of distinct orbits $\Orb[\Phi]{j}$ in $\Events{\jointvar}$. For the Web inflation in particular, $M_{\Phi}$ is $450 \times 358,120$. Using the symmetric incidence matrix and linear programing methods, an infeasibility certificate was found that is capable of witnessing the Fritz distribution. The corresponding deflated inequality is presented in \cref{sec:found_inequalities} as \cref{eq:symmetric_web_inequality}.

    \nocite{apsrev41Control}
    \bibliography{references}

\end{document}